\documentclass[%
 reprint,
 superscriptaddress,
 amsmath,amssymb,
 aps,
pra,
]{revtex4-2}

\usepackage{graphicx}
\usepackage{dcolumn}
\usepackage{bm}
\usepackage[separate-uncertainty=true,multi-part-units=single]{siunitx} 

\begin{document}

\title{Modelling Active Non-Markovian Oscillations}

\author{G. Tucci}
 \email{gtucci@sissa.it}
\affiliation{SISSA --- International School for Advanced Studies and INFN, via Bonomea 265, 34136 Trieste, Italy}

\author{\'E. Rold\'an}
    \email{edgar@ictp.it}
\affiliation{%
    ICTP --- The Abdus Salam International Centre for Theoretical Physics, Strada Costiera 11, 34151  Trieste, Italy
}%

\author{A. Gambassi}
\email{gambassi@sissa.it}
\affiliation{SISSA --- International School for Advanced Studies and INFN, via Bonomea 265, 34136 Trieste, Italy}

\author{R. Belousov}
    \email{belousov.roman@gmail.com}
    \homepage{https://belousov.tel}
\affiliation{%
    ICTP --- The Abdus Salam International Centre for Theoretical Physics, Strada Costiera 11, 34151  Trieste, Italy
}%
\affiliation{
    EMBL --- European Molecular Biology Laboratory, Meyerhofstr.~1, 69117, Heidelberg, Germany
}

\author{F. Berger}
\affiliation{Cell Biology, Neurobiology and Biophysics, Department of Biology, Faculty of Science, Utrecht University, 3584 CH, Utrecht, The Netherlands}

\author{R. G. Alonso}
\affiliation{Howard Hughes Medical Institute and Laboratory of Sensory Neuroscience, The Rockefeller University, 1230 York Avenue, New York, NY 10065, USA}

\author{A. J. Hudspeth}
\affiliation{Howard Hughes Medical Institute and Laboratory of Sensory Neuroscience, The Rockefeller University, 1230 York Avenue, New York, NY 10065, USA}


\begin{abstract}
Modelling noisy oscillations of active systems is one of the current challenges in physics and biology.
Because the physical mechanisms of such  processes are often difficult to identify, we propose a  linear stochastic model driven by a non-Markovian bistable noise that is capable of generating self-sustained periodic  oscillation. We derive analytical predictions  for most relevant dynamical and thermodynamic properties of the model. This minimal model turns out to  describe accurately bistable-like oscillatory motion of hair bundles in  bullfrog sacculus, extracted from experimental data. Based on and in agreement with these data, we estimate the power required to sustain such active oscillations to be of the order of one hundred $k_B T$ per oscillation cycle.
\end{abstract}

\maketitle

Most non-equilibrium systems \textit{actively} sustain their dynamics by dissipating energy into their environment and by producing entropy, as observed in several branches of natural sciences \cite{kruse2005oscillations,Toner,rama,hudspethIntegratingActiveProcess2014,prost2015active,Cates,bechinger2016active,FODOR,das2020introduction,PhysRevE.97.032604,shreshtha2019thermodynamic,cerasoli2022spectral}. 
Important examples are \textit{active oscillators}, the effective mesoscopic degrees of freedom of which are described by oscillating variables. In nature these oscillators drive climate changes, sustain heart beat, facilitate vocal and auditory systems, and support neural signaling and circadian rhythms \cite{kruse2005oscillations,hudspethIntegratingActiveProcess2014,Martin,Martin4533,alonsoMotorControlSound2014,Alejo,betaIntracellularOscillationsWaves2017,buzsakiNeuronalOscillationsCortical2004,cherevkoAnalysisSolutionsBehaviour2017a,cherevkoRelaxationOscillationModel2016a,fitzhugh1961impulses,mindlinNonlinearDynamicsStudy2017,mirolloSynchronizationPulseCoupledBiological1990,nagumoActivePulseTransmission1962,nomuraBonhoeffervanPolOscillator1993,oatesPatterningEmbryosOscillations2012,roennebergModellingBiologicalRhythms2008,rompalaDynamicsThreeCoupled2007,vanderpolBiologicalRhythmsConsidered1940,vanderpolLXXIIHeartbeatConsidered1928,vandijkAmplitudeFrequencyFluctuations1990,vettorettiFastPhysicsSlow2018,belousov2020volterra}.   Here we focus on those responsible for mechanoelectrical transduction in the bullfrog's sacculus, for which experimental data are available~\cite{hudspethIntegratingActiveProcess2014,Martin,Martin4533}.

    Active oscillatory motion is often interpreted as relaxation oscillations or noisy bistable oscillations, which can be modeled by nonlinear stochastic Van der Pol and Duffing equations \cite{hudspethIntegratingActiveProcess2014,Martin,Martin4533,alonsoMotorControlSound2014,Alejo,betaIntracellularOscillationsWaves2017,buzsakiNeuronalOscillationsCortical2004,cherevkoAnalysisSolutionsBehaviour2017a,cherevkoRelaxationOscillationModel2016a,fitzhugh1961impulses,mindlinNonlinearDynamicsStudy2017,mirolloSynchronizationPulseCoupledBiological1990,nagumoActivePulseTransmission1962,nomuraBonhoeffervanPolOscillator1993,oatesPatterningEmbryosOscillations2012,roennebergModellingBiologicalRhythms2008,rompalaDynamicsThreeCoupled2007,vanderpolBiologicalRhythmsConsidered1940,vanderpolLXXIIHeartbeatConsidered1928,vandijkAmplitudeFrequencyFluctuations1990,vettorettiFastPhysicsSlow2018,belousov2020volterra,belousovVolterraseriesApproachStochastic2019,maoileidighDiverseEffectsMechanical2012}, respectively. These two distinct  dynamical regimes are not always easy to distinguish in experiments. An alternative way to construct a system displaying bistable oscillations consists in letting one of its degrees of freedom to be a two-state stochastic process such as  telegraph noise \cite{buceta2001stationary,muller2015statistics,huber2003dynamics,yuzhelevski2000random,mankin2007noise,kurzynski2008statistical,gurvitz2016temporal,aharony2019telegraph,wittrock2021flicker}. 

In this work,  we propose a stochastic linear model for self-sustained, active, bistable oscillations. The model generalizes the Ornstein-Uhlenbeck process by allowing the equilibrium position (the center of the harmonic potential) to be determined by a dichotomous non-Markovian noise. 
Notably, depending on the distributions of the waiting times, the model can reproduce a wide variety of bistable oscillations including Markovian and non-Markovian switching processes.  We obtain exact analytical expressions for several  dynamical and thermodynamic quantities characterizing the nonequilibrium nature of the system.  
As a relevant application, we use our model to reproduce  recordings of the spontaneous motion in bullfrog hair bundles and  estimate the dissipated power, which is experimentally  inaccessible but crucial for interpreting the energetics  of system. \smallskip

\begin{figure*}[t!]
\centering
\includegraphics[width = \linewidth]{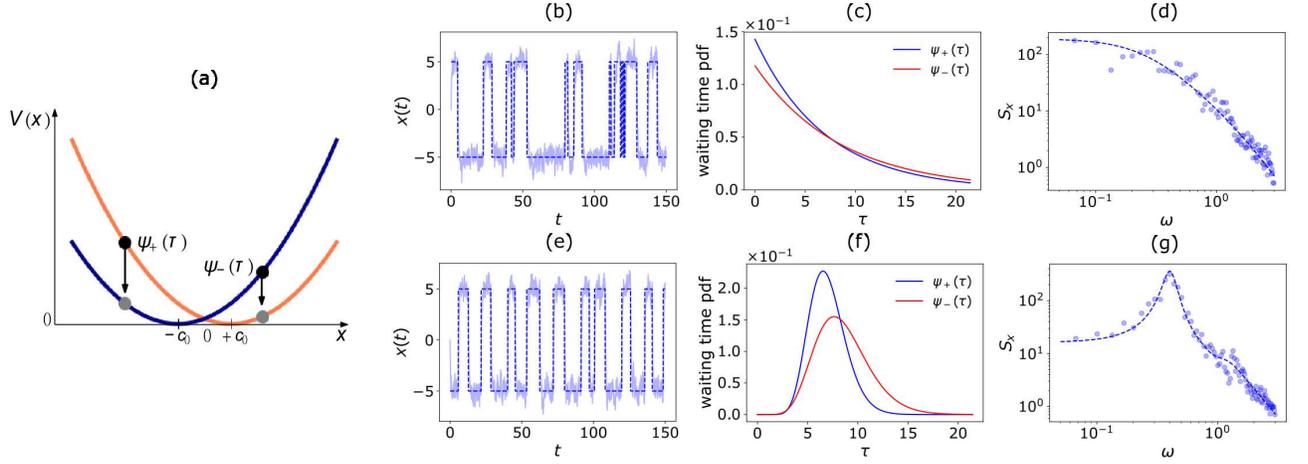}
\caption{Panel (a) represents the two possible switching mechanisms controlling the dynamics described in Eq. \eqref{eq:langevin}: after a time $\tau$ drawn from the distribution $\psi_\pm(\tau)$ the center $c$ of a harmonic potential $V(x)=(\kappa/2) (x-c)^2$ switches from $\pm c_0$ to $\mp c_0$.  Second column: realizations of the stochastic driving $c(t)$ (dashed blue line)  and of the process $x(t)$  (solid blue line) obtained from a numerical simulation of Eq.~\eqref{eq:langevin}, for (b) the exponential and (e) the gamma waiting-time distributions plotted, respectively, in panels (c) and (f) of the second column. 
%
%
%
In particular, the exponential distributions have rates $r_+=1/7$, and $r_-=2/17$, whereas the gamma distributions (see the main text) have shape parameters $k_+=15$, $k_-=10$, and scale parameters $\theta_+=7/15$, $\theta_-=17/20$.
Last column: Power spectral density $S_x$ (symbols) of $x(t)$ on the doubly  logarithmic scale, obtained for two time series of total duration $t=1.5\times 10^3$ with the same parameters as those in panels (b) and (d).  The dashed lines are given by Eq.~\eqref{eq:psd}. 
The dynamics was simulations with $D=1$, $c_0=5$, $\nu=2.5$, and a time step $\Delta t=10^{-3}$. 
}\label{fig:traj}
\end{figure*}

\textit{Model}.---We consider an Ornstein-Uhlenbeck process $x(t)$ with time-dependent  center $c(t)$  described by the  stochastic differential equation
\begin{equation}\label{eq:langevin}
\gamma\,\dot{x}(t)=-\kappa [x(t)-c(t)]+\xi(t).
\end{equation}
Here $\kappa$ is the stiffness of the harmonic potential $V(x,c)=\kappa(x-c)^2/2$, $\gamma$ is the effective friction coefficient, and  
$\xi(t)$ is a Gaussian white noise with zero mean $\langle \xi(t) \rangle=0$ and autocorrelation $\langle \xi(t_1) \xi(t_2) \rangle=2\gamma^2 D\delta(t_1-t_2)$, where the effective diffusion coefficient $D=k_B T/\gamma$ is related to the temperature through the Einstein relation.
The center $c(t)$ is a dichotomous process taking the values $\pm c_0$,  with $c_0\geq 0$, and changing sign at stochastic intervals. 
We denote by $\psi_\pm(\tau)$  
the distribution of the waiting time spent in $\pm c_0$ 
before switching sign; we refer to Fig. \ref{fig:traj}a for an illustration. 
The  relevant timescales of the dynamics are  the two mean waiting times   $\langle \tau \rangle_\pm\equiv \int_0^\infty\mathrm{d}\tau\,\tau\,\psi_\pm(\tau)$ and the relaxation time $\tau_\nu=\nu^{-1}$ in the harmonic potential, where $\nu = \kappa/\gamma $.
Note that $c(t)$ is a non-Markovian process unless the two waiting-time distributions are exponential $\psi_{\pm }(\tau)=e^{-\tau/\langle \tau \rangle_\pm}/\langle \tau \rangle_\pm$. In this case, $c(t)$ 
corresponds to the so-called (Markovian) telegraph noise~\cite{kac1974stochastic}.

%
%
%
Figure~\ref{fig:traj} shows representative trajectories of the process  $x(t)$ for various choices of the waiting-time distributions $\psi_{\pm}(\tau)$.   Panel (b) refers to the exponentially-distributed waiting times reported in panel (c), with the Lorentzian power spectrum (see further below) shown in panel (d). 
Panel (e), instead, shows a realization of the process  $x(t)$ for the gamma-distributed waiting times  reported in panel (f), which is characterized by fast jumps between the two (almost) equilibrium states. The power spectrum of the process, shown in panel (g), features a pronounced peak at the typical frequency of the coherent oscillations.  As we will show below, the interplay between $\langle\tau\rangle_{\pm}$ and $\tau_\nu$ determines which type of  stationary dynamics, either monostable or bistable, emerges from the fluctuations of the system.

\begin{figure}
\includegraphics[width = 0.85\columnwidth]{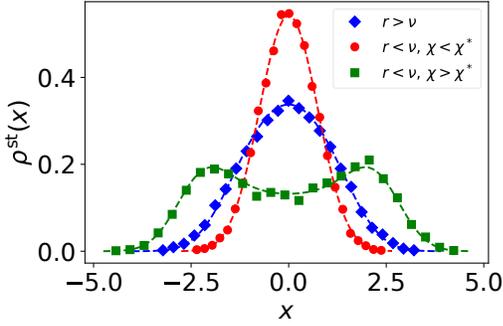}
\caption{Stationary probability density $\rho^{\rm st}(x)$ for symmetric exponentially-distributed waiting times: numerical simulations (symbols) are compared with the analytical solution in Eq.~\eqref{eq:rhosol} (dashed lines). The three cases correspond to fixed values of $D=1$ and $\nu=2.5$, but various values of  $r$ and $c_0$: 
(blue)  $c_0=2$ and $r=5>\nu$; 
(red) $c_0=0.5$ and $r=1.25<\nu$ for which $\chi \simeq 0.31$;
(green) $c_0=2.5$ and $r=1.25$, for which $\chi \simeq 7.8$. 
In the latter two cases $\zeta=1/2$ corresponding to the critical value $\chi^*(\zeta=1/2)\simeq 1.58$ (see main text). 
The  numerical estimates of $\rho^{\rm st}(x)$ are obtained from $N=10^4$ simulations of Eq.~\eqref{eq:langevin} using  Euler's numerical integration method with time step $\Delta t=5\times 10^{-3}$. }\label{fig:rho}
\end{figure}

\textit{Dynamics}.---We encode  the state of the system at time~$t$ by the couple of stochastic  variables $\left(x(t),\sigma(t)\right)$, where  $\sigma(t)= c(t)/c_0 = \pm 1$ 
is the sign of  $c(t)$. 
A quantity of interest is the joint probability density $\rho_{\sigma} (x,t|x_0)$  for the system  to be in the state $(x,\sigma)$ at time $t$ given that its initial state was $x(0)=x_0$.  Its normalization requires 
$\sum_{\sigma} \int {\rm d}x \,\rho_{\sigma} (x,t|x_0) = 1$ for all times $t\geq 0$. 
We derive a renewal equation for $\rho_{\sigma} (x,t|x_0)$ (see Appendix \ref{App:probability density})
in terms of the waiting-time distributions $\psi_\sigma(\tau)$ and of the probability density 
$G_\sigma^{(0)}(x,t|x_0)$. The latter  is given by the probability density to be in $x$ at time $t$  for an Ornstein-Uhlenbeck process with fixed center $c(t)=\sigma c_0$. From the  formal expression of $\rho_{\sigma} (x,t|x_0)$, we determine the analytical expressions of
the Laplace transform of the first and second moments of $x(t)$ for generic waiting-time distributions $\psi_\sigma(\tau)$.

Because switches break detailed balance, the system reaches a nonequilibrium stationary state at long times. 
For exponentially-distributed waiting times with rates $r_\sigma=1/\langle\tau \rangle_\sigma$, we find 
an explicit expression of the stationary distributions  $\rho_\sigma^{\rm st}(x)=\lim_{t\to\infty}\rho_\sigma(x,t|x_0)$.  In this case, the finite-time densities $\rho_\sigma(x,t|x_0)$ satisfy 
Fokker-Planck equations with source terms
\begin{equation}\label{eq:roexp}
\begin{aligned}
\partial_t\rho_\sigma(x,t|x_0)=&-\partial_x J_\sigma(x,t)\\
&+r_{-\sigma}\rho_{-\sigma}(x,t|x_0)-r_\sigma\rho_{\sigma}(x,t|x_0),
\end{aligned}
\end{equation}
where $J_\sigma(x,t)=-\left[\nu(x-\sigma c_0)+D\partial_x\right]\rho_\sigma(x,t|x_0)$ is the spatial probability current associated to particles in the state $\sigma$ at time $t$. 
The stationary solutions $\rho^{\rm st}_\sigma(x)$ of  Eq.~\eqref{eq:roexp}
are then given by
\begin{equation}\label{eq:rhosol}
\rho^{\rm st}_\sigma(x)=\frac{\mathcal{N}}{2}\int_{-1}^{+1}\mathrm{d}z \,\rho_{\rm G}(x-c_0z) (1-\sigma z)^{r_\sigma/\nu-1}(1+\sigma z)^{r_{-\sigma}/\nu},
\end{equation}
where we introduce the Gaussian  distribution  ${\rho_{\rm G}(x)\equiv\exp[-x^2\nu/(2D)]/\sqrt{2\pi D/\nu}}$. The constant $\mathcal{N}^{-1}\equiv\nu\sum_\sigma{{}_2F_1(1,1-r_{-\sigma}/\nu,1+r_\sigma/\nu,-1)/r_\sigma}$ enforces normalization of $\rho^{\rm st}_\sigma(x)$, where ${}_2F_1$ is the hypergeometric function.  
Similar results were recently reported for 
run-and-tumble particles~\cite{Garcia,Confpot}. 
The total stationary density $\rho^{\rm st}(x)=\rho^{\rm st}_+(x)+\rho^{\rm st}_-(x)$  can be 
either unimodal or bimodal, 
depending on the values of the parameters of the model, as shown in Fig.~\ref{fig:rho} numerically and analytically using Eq.~\eqref{eq:rhosol}.  
In particular, bistability emerges whenever the relaxation is fast enough with respect to the switching frequency. For symmetric and exponentially-distributed waiting times, i.e. $r_\sigma=r$, we can  characterize the transition from unimodal to bimodal analytically, exploiting the fact that $\rho^{\rm st}(x)$ is unimodal if it displays a maximum at $x=0$, and  bimodal otherwise. The transition is controlled 
by the dimensionless parameters $\zeta\equiv r/\nu$, describing the interplay between relaxation and switching, and $\chi\equiv c_0^2\nu/(2D)$, 
which quantifies how much the two centers $\pm c_0$ are distinguishable with respect to the amplitude of thermal fluctuations. 
%
%
We find that for fast switching $r\geq\nu$ ($\zeta\geq 1$) the stationary distribution is always unimodal as shown by the blue curve and data points in Fig.~\ref{fig:rho}, whereas for slow switching $r<\nu$ ($\zeta<1$) $\rho^{\rm st}(x)$ can display both monostability and bistability. In particular, the dynamics is monostable for $\chi\leq\chi^*(\zeta)$, as shown in red in Fig.~\ref{fig:rho}, and bistable for $\chi>\chi^*(\zeta)$, shown in green. The critical value $\chi^*(\zeta)$ depends solely on $\zeta$ (see Appendix~\ref{App:Markovian}). 

Another relevant quantity that characterizes the dynamics of the system is the long-time correlator  ${C_x(t)\equiv\lim_{\tau\to\infty}\langle x(t+\tau)x(\tau)\rangle}$. Its Fourier transform is the power spectral density  $S_x(\omega)= \hat{C}_x(\omega)=\langle|\hat{x}(\omega)|^2\rangle$, where we  use the convention  $\hat{f}(\omega)\equiv\int_{-\infty}^{+\infty}\mathrm{d}t\,e^{-i\omega t}f(t)$ for the Fourier transform $\hat{f}$ of a function $f$.
Because the noise terms $\xi$ and $c$ in Eq.~\eqref{eq:langevin}  are independent, it follows  that
\begin{equation}\label{eq:Sx}
S_x(\omega)=\frac{2D+\nu^2 S_c(\omega)}{\nu^2+\omega^2},
\end{equation}
where $S_c(\omega)=\langle|\hat{c}(\omega)|^2\rangle$ is the power spectrum of $c(t)$. 
For generic non-Markovian  $c(t)$,  calculating  
$S_c(\omega)$ requires the knowledge of its stationary two-time statistics 
derived in Appendix \ref{App:B}. 
In particular, the key quantity is the 
Laplace transform $\widetilde{C}_c(s)$ of the long-time $c$-correlator $C_c(t)$, defined as above, which is given by
\begin{equation}\label{eq:Cs}
\widetilde{C}_c(s)=c_0^2\left[\frac{1}{s}-\frac{2}{\langle \tau\rangle}\frac{\widetilde{\Psi}_-(s)\widetilde{\Psi}_+(s)}{1-\widetilde{\psi}_-(s)\widetilde{\psi}_+(s)}\right].
\end{equation}
Here we define the Laplace transform of $f$ as $\widetilde{f}(s)\equiv \int_0^\infty\mathrm{d}t\,e^{-st}f(s)$, thus $\widetilde{\psi}_\sigma(s)$ and $\widetilde{\Psi}_\sigma(s)=[1-\widetilde{\psi}_\sigma(s)]/s$ are, respectively, the transforms of the waiting-time distribution $\psi_\sigma(t)$ and of its cumulative $\Psi_\sigma(t)=\int_t^\infty\mathrm{d}\tau\,\psi_\sigma(\tau)$ whereas $\langle \tau\rangle\equiv(\langle \tau\rangle_++\langle \tau\rangle_-)/2$ is the average half-period of the oscillations.
The analyticity of $\widetilde{C}_c(s)$ on the imaginary axis implies that $S_c(\omega)=\widetilde{C}_c(i\omega)+\widetilde{C}_c(-i\omega)$.

\begin{figure}[h]
\includegraphics[width = 0.95\linewidth]{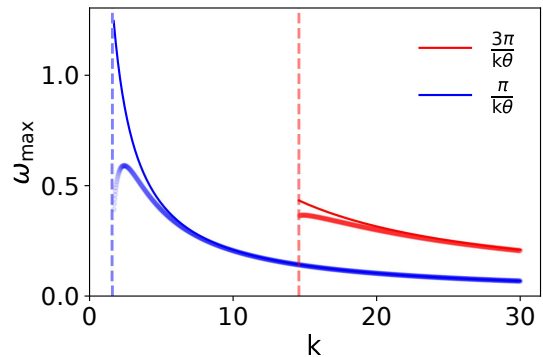}
\caption{Frequencies of the first (blue) and second (red) peaks of the power spectrum $S_x(\omega)$ in Eq.~\eqref{eq:Sx}, as functions of $k$, for $\theta=1.5$, $D=0.5$, $c_0=1$, and $\nu=2.5$. As $k$ increases above $\simeq 1.5$ (dashed blue vertical line) a first local maximum appears in $S_x(\omega)$ at a typical frequency (blue symbols), well approximated by the blue solid line.  As $k$ exceeds  $\simeq 14.6$ (dashed red vertical line), a second peak appears at a typical frequency (red symbols), well approximated by the red solid line.
}
\label{fig:omegamax}
\end{figure}

Recent works \cite{tu2008nonequilibrium,skinner2021estimating} revealed that non-monotonic waiting-time distributions 
often emerge from 
underlying nonequilibrium stationary process.
These features may be described by gamma-distributed waiting times ${\psi_\sigma(\tau)=\left(\theta_\sigma^{k_\sigma}\Gamma(k_\sigma)\right)^{-1}\tau^{k_\sigma-1}e^{-\tau/\theta_\sigma}}$, with average  $\langle \tau\rangle_\sigma=k_\sigma\theta_\sigma$ and 
Laplace transforms $\widetilde{\psi}_\sigma(s)=(1+s\theta_\sigma)^{-k_\sigma}$. For this example, the power spectrum $S_c(\omega)$ reads
\begin{equation}
\begin{aligned}
&S_c(\omega)=\frac{4c_0^2}{\langle \tau\rangle \omega^2}\times\\
&\frac{(R_+ R_- )^2-1+(1-R_-^2)R_+\cos\phi_++(1-R_+^2)R_-\cos\phi_-}{(R_+ R_- )^2+1-2R_+ R_- \cos(\phi_++\phi_-)},\label{eq:psd}
\end{aligned}
\end{equation}
where we define $\phi_\sigma(\omega)\equiv k_\sigma\arctan\left(\omega\theta_\sigma\right)$, and  $R_\sigma(\omega)\equiv(1+\omega^2\theta^2_\sigma)^{k_\sigma/2}$. 
Equation~\eqref{eq:psd} agrees   
with the numerical estimates of the power spectrum for both exponentially- and gamma-distributed waiting times, as shown in panels (c) and (f), respectively, of Fig.~\ref{fig:traj}.
We find that the power spectrum $S_x(\omega)$ displays a peak at a 
frequency $\omega_{\mathrm{max}}$ for sufficiently large values of $k$, which depend on the choice of parameters. Moreover, for large values of $k$,  the spectrum may display additional peaks close to the integer multiples of  $\omega_{\mathrm{max}}$.
For symmetric gamma-distributed waiting  times, Fig.~\ref{fig:omegamax} shows the frequencies corresponding to the first two peaks of $S_x(\omega)$ as a function of the shape parameter $k$. We note that the second peak appears for $k\gtrsim 14.6$ at frequency $\simeq 3\omega_{\rm max}$, which results from the fact that, upon increasing $k$, $c(t)$ increasingly resembles a deterministic symmetric square wave whose Fourier spectrum has only odd harmonics.
%
%

\textit{Stochastic thermodynamics}.---To characterize the thermodynamics of the active mechanism driving the oscillations, we evaluate the statistics of the work. 
The stochastic work $\delta W(t)$ \cite{sekimoto1998langevin} done on the system in the time interval $[t,t+\mathrm{d}t]$ is given by $\delta W(t)=(\partial  V/\partial c)\circ \mathrm{d}c(t)=-\kappa\,x(t)\circ \mathrm{d}c(t)$, 
where $\circ$ denotes the Stratonovich product, and the second equality follows from $c^2(t)=c_0^2$.
Note that $\delta W(t)$ is non-zero only when a switch occurs at time $t$: energy is injected into (extracted from) the system, i.e., $\delta W(t)>0$ ($\delta W(t)<0$), when $x(t)\mathrm{d}c(t)<0$  ($x(t)\mathrm{d}c(t)>0$).
From the analytical expression of the first moment of $x(t)$ at a switch, we derive the exact expression of the stationary average power $\langle\dot{W}\rangle=\lim_{t\to\infty}\langle \delta W(t)\rangle/\mathrm{d}t$:
\begin{equation}\label{eq:dissipation}
\displaystyle\langle \dot{W}\rangle=\frac{2\kappa c_0^2}{\langle \tau\rangle}\frac{[1-\widetilde{\psi}_+(\nu)][1-\widetilde{\psi}_-(\nu)]}{1-\widetilde{\psi}_+(\nu)\widetilde{\psi}_-(\nu)},
\end{equation}
which holds for \textit{arbitrary} waiting-time distributions~$\psi_\sigma(\tau)$.
%
%
The average stationary power is always positive in agreement with the second law:  
$\langle\dot{W}\rangle=T \langle\dot{S}_{\rm tot}\rangle\ge 0$, where $\langle\dot{S}_{\rm tot}\rangle$ is the rate of entropy production. 
Moreover, Eq.~\eqref{eq:dissipation} implies the upper bound $\langle\dot{W}\rangle \le 2\kappa c_0^2/\langle \tau\rangle$, 
which is saturated in the limit of infinitely fast relaxation time  ($\nu\rightarrow\infty$). 
The upper bound $2\kappa c_0^2/\langle\tau\rangle$ is the ratio between the characteristic energy $V_0=\kappa (2c_0)^2/2$
that $x(t)$ fluctuating around the minimum of one potential acquires in the other potential immediately after the switch
and  the average time between successive switches $\langle \tau\rangle$. 
For $\nu\rightarrow \infty$, this $V_0$ is indeed the energy injected in the system at a switch. 
Furthermore, we derive in Appendix \ref{App:C} 
exact expressions of the average work $\langle W(t)\rangle=\int_0^t \langle\delta W(\tau)\rangle$ and of 
$\langle W^2(t)\rangle$.

\textit{Experimental application}.---An example of a biological process displaying active oscillations is the spontaneous motion of hair bundles from a bullfrog's ear \cite{hudspethIntegratingActiveProcess2014,Martin,Martin4533}. The hair bundle is an organelle formed by a cohesive tuft of cylindrical stereocillia that protrude from the apical surface of the namesake hair cells. This receptor cells transduces a mechanical stimulus, such as a sound wave, into a neural signal and thus facilitates hearing and other sensory processes in  vertebrates. The oscillatory motion of a hair bundle is powered by an active process, which is essential for the organelle's sensory function, and results in the violation of the fluctuation-dissipation theorem \cite{Rol}.

Several stochastic models have been proposed for the time series of hair bundles 
\cite{choeModelAmplificationHairbundle1998,Martin4533,tinevezUnifyingVariousIncarnations2007,reichenbachPhysicsHearingFluid2014,Martin,vilfanTwoAdaptationProcesses}. All these models, which can be reduced to the family of Duffing -- Van der Pol oscillators \cite{maoileidighDiverseEffectsMechanical2012,belousovVolterraseriesApproachStochastic2019,belousov2020volterra,Martin}, rely on nonlinear equations of motion with hidden degrees of freedom of diverse origins. Under various conditions such a system can describe both bistable and limit-cycle regimes of oscillatory motion, which are often not easy to distinguish.

\begin{figure}[h]
    \centering
    \includegraphics[width=0.8\columnwidth]{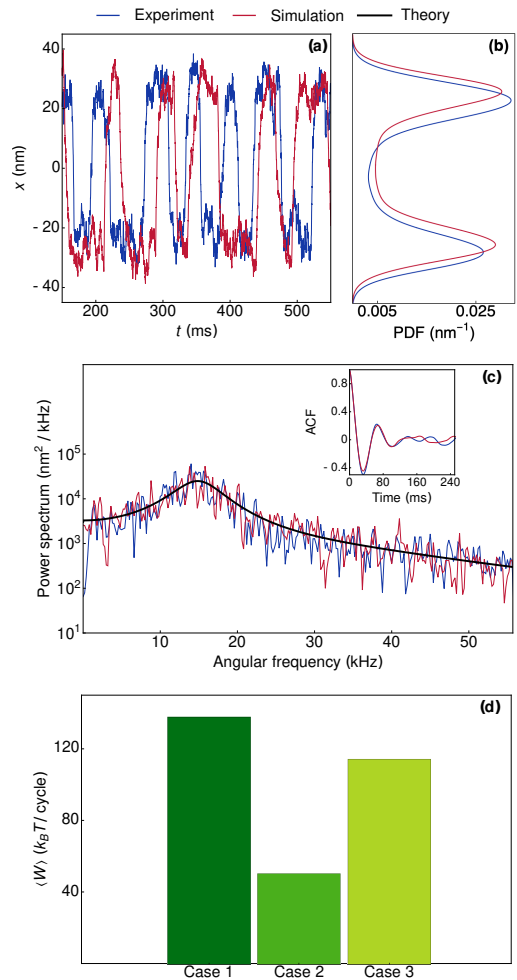}
    \caption{
        Oscillations of a hair bundle's tip $x(t)$ modelled by Eq.~\eqref{eq:langevin}: experimental observations and simulations with inferred parameter values. (a) Example segments of experimental and simulated time series.  (b) Probability density of $x(t)$. (c) Power spectrum of $x(t)$ with its autocorrelation function shown in the inset. (d) Energy dissipated by hair-bundles per one cycle in three experimental cases, see Table~\ref{tab:1} in Appendix~\ref{app:fit}. Only data of the experimental case 1 are shown in panels (a)--(c). Data for all the three cases are reported in Appendix~\ref{app:fit}.
    }
    \label{fig:summary}
\end{figure}
In typical experiments, oscillating hair bundles display a great variety of different non-linear oscillations \cite{Martin4533}. We applied our   theoretical model to symmetric bi-stable oscillations and therefore we specifically select appropriate traces from our experimental recordings. These measurements were performed on a dissected mechanosensitive epithelium of a bullfrog's sacculus, as described previously \cite{alonso2020fast,azimzadeh2017physiological}.
In an experiment, we mounted the mechanosensitive tissue in a two-compartment chamber, such that the hair cells were exposed to two different ionic solutions on their apical and basal side. This setup mimicked the physiological condition in which hair cells
operate in the inner ear and evoked spontaneous oscillations of the hair bundles. To better resolve the movement of the oscillating hair bundle, we attached a glass fiber to the bundle's tip and projected the shadow onto a photodiode \cite{azimzadeh2018thermal}. This calibrated signal of the photodiode reported the position of the oscillating bundle as a function of time (blue line in Fig. \ref{fig:summary}a).

As reported below, the linear model proposed in this letter is also capable to account for the basic features of the hair-bundle motion, which are common to simple active oscillators, see  Fig.~\ref{fig:summary}a--c. To make contact with the experimental data we apply a simulation-based inference approach \cite{greenberg2019,papamakarios2016,lueckmann2017,greenberg2019}
(Appendix~\ref{app:fit}) to determine values of the unknown parameters in Eq.~\eqref{eq:langevin} for a selection of three experimental cases in which we observed simple symmetric oscillations of $x(t)$. Our model reproduces well the pattern of the hair-bundle motion as shown in Fig.~\ref{fig:summary}a. The simulated time-series of $x(t)$ also quantitatively  match the probability density and time-frequency statistics of experimental measurements, see  Fig.~\ref{fig:summary}b-c.
Using the exact analytical predictions obtained for the equation of motion~\eqref{eq:langevin}, we can estimate the average power dissipated by the active process that drives the hair-bundle oscillations in the bullfrog's ear. Its value Fig.~\ref{fig:summary}d,  $\langle\dot{W}\rangle\sim 100 \,k_{\rm B}T$/cycle, is of the same order of magnitude as estimates of the heat dissipation rate in hair-bundle spontaneous fluctuations~\cite{Rol}, and of the  viscous energy dissipation under weak, external periodic stimulation~\cite{martin1999active}. 
Assuming that  active oscillations result from ATP hydrolysis by myosin motors with  a free energy change of $\sim 10k_{\rm B}T$ per molecule~\cite{hudspeth1994pulling}, we estimate about ten ATP molecules are required to fuel  a single oscillation cycle of the hair bundle.

\textit{Discussion}.---In this work, we have introduced an exactly-solvable stochastic model describing the dynamics of non-Markovian active oscillators. This system displays key dynamical features of active oscillators: transition from a monostable to a  bistable regime,  sharp power spectra, broken detailed balance, and heat dissipation.  We have also  generalized the theoretical analysis presented here to accommodate asymmetric waiting-time distributions of the underlying noise (see Appendix \ref{App:probability density}), as observed in many biological processes~\cite{skinner2021estimating}.  We have also shown that our linear, non-Markovian model   reproduces with high accuracy the probability density and power spectrum of several experimental recordings  from the top of the bullfrog's saccular hair bundle. 
Fitting the data to the model, we have calculated that the power consumption by the hair bundle during its spontaneous motion requires the consumption of at least ten ATP molecules per oscillation cycle. We expect that our model could be applied to decipher the energetics of other relevant active oscillations observed in living systems, such as confined cell migration~\cite{bruckner2021learning}, neuronal networks~\cite{nadkarni2003spontaneous}, and actomyosin gels~\cite{placcais2009spontaneous}.

\providecommand{\noopsort}[1]{}\providecommand{\singleletter}[1]{#1}

\newpage

\onecolumngrid

\appendix
\vskip3em
\begin{center}\small\textbf{APPENDIX}\end{center}
\section{Probability density of the process}\label{App:probability density}

In this Appendix, we address the calculation of $G(x,\sigma,t|x_0,\sigma _0)$, the probability density of finding the system in the state $(x,\sigma )$ at time $t$, given that it starts from $(x_0,\sigma_0 )$. We can express $G(x,\sigma,t|x_0,\sigma_0)$ by means of renewal theory in terms of the probability density $G^{(0)}_\sigma(x,t|x_0)$ of the switch-free dynamics, the waiting-time distribution $\psi_\sigma(t)$, and its cumulative
\begin{equation}\label{eq:Psi}
\Psi_\sigma(t)\equiv\int_t^\infty\mathrm{d}\tau\,\psi_\sigma(\tau).
\end{equation}

Before proceeding to the computation of $G(x,\sigma,t|x_0,\sigma_0)$, we introduce $G_S(x,\sigma,t|x_0,\sigma_0 )$, the probability density associated with trajectories that start at $(x_0,\sigma_0)$ and reach the state $(x,\sigma )$ at time $t+\mathrm{d}t$, conditioned on the fact that at least one switch occurs in the time interval $(0,t)$ with a last switch at time $t$. The expression of $G_S(x,\sigma,t|x_0,\sigma_0)$ depends on the initial and final configuration, and we calculate it by considering all the possible number of switches occurring within the time interval $(0,t)$:

\begin{equation}\label{eq:GS}
\begin{aligned}
G_S(x,\sigma,t|x_0,\sigma )=&\sum_{n=1}^\infty\int_0^t\mathrm{d}\tau_1\int_{-\infty}^{+\infty}\mathrm{d}z_1\,G_\sigma(z_1,\tau_1|x_0)\int_0^{t-\tau_1}\mathrm{d}\tau_2\int_{-\infty}^{+\infty}\mathrm{d}z_2\,G_{-\sigma}(z_1,\tau_2|z_1)\\
&+\cdots \int_0^{t-\sum_{k=1}^{2n-2}\tau_k}\mathrm{d}\tau_{2n-1}\int_{-\infty}^{+\infty}\mathrm{d}z_{2n-1}\,G_\sigma(z_{2n-1},\tau_{2n-1}|z_{2n-2})\,G_{-\sigma}\left(x,t-\sum_{k=1}^{2n-1}\tau_k\Bigg|z_{2n-1}\right)\\
=&\int_{-\infty}^{+\infty}\mathrm{d}y\int_0^t\mathrm{d}\tau\,G_{-\sigma}(x,t-\tau|y)G_S(y,-\sigma,\tau|x_0,\sigma ),
\end{aligned}
\end{equation}
and
\begin{equation}\label{eq:GS2}
    G_S(x,-\sigma,t|x_0,\sigma )=G_\sigma(x,t|x_0)+\int_{-\infty}^{+\infty}\mathrm{d}y\int_0^t\mathrm{d}\tau\,G_\sigma(x,t-\tau|y)G_S(y,\sigma,\tau|x_0,\sigma ),
\end{equation}
where we define the switch-free total probability density 

\begin{equation}
G_\sigma(x,t|x_0)\equiv \psi_\sigma(t)\,G_\sigma^{(0)}(x,t|x_0),
\end{equation}
and we identify the initial state $(x_0,\sigma_0)$ as a state displaying a switch.

The sum over $n$ in the expression of  $G_S(x,\sigma,t|x_0,\sigma )$ corresponds to all possible number of switches in $(0,t)$ with same initial and final state $\sigma$: this fixes an even number of switches, including the last one at time $t$. Then, the evolution from the initial to final state is given by the alternation of ``bare" probability density $G_\sigma(x,t|x_0)$, describing the free dynamics between two switches. Similarly, the probability density $G_S(x,-\sigma,t|x_0,\sigma )$ is compatible with trajectories displaying an odd number of switches: the first contribution $G_\sigma(x,t|x_0)$ accounts for one single switch at time $t$, the second with any odd number of switches larger than one.
Given that the waiting-time distribution associated the last switch is not integrated over time, we can immediately deduce that $G_S(x,\sigma,t|x_0,\sigma_0)$ is a probability density also with respect to $t.$ Indeed, we interpret $G_S(x,\pm \sigma,t|x_0,\sigma )\mathrm{d}x\mathrm{d}t$ to be the probability to switch within the time interval $(t,t+dt)$ to the state $\pm \sigma$ to a position in $(x,x+\mathrm{d}x)$, given the initial state $(x_0,\sigma )$.
Integrating Eqs. \eqref{eq:GS} with respect the final position $x$ we get
\begin{equation}
P_S(\pm \sigma,t|\sigma )\equiv \int\mathrm{d}x\,G_S(x,\pm \sigma,t|x_0, \sigma ),
\end{equation}
i.e., the probability density to be initially in the state $\sigma$ and to end up in $\pm \sigma$ after a switch at a time in $(t,t+\mathrm{d}t)$:

\begin{equation}\label{eq:PSt}
\begin{aligned}
P_S(\sigma,t|\sigma )=&\sum_{n=1}^\infty\int_0^t\mathrm{d}\tau_1\,\psi_\sigma(\tau_1)\int_0^{t-\tau_1}\mathrm{d}\tau_2\psi_{-\sigma}(\tau_2)\cdots \int_0^{t-\sum_{k=1}^{2n-2}\tau_k}\mathrm{d}\tau_{2n-1}\psi_\sigma(\tau_{2n-1})\,\psi_{-\sigma}\left(t-\sum_{k=1}^{2n-1}\tau_k\right),\\
P_S(-\sigma,t|\sigma )=&\,\psi_\sigma(t)+\int_0^t\mathrm{d}\tau\,\psi_\sigma(t-\tau)P_S(\sigma,\tau|\sigma ),
\end{aligned}
\end{equation}
which follows from the normalization $\int_{-\infty}^{+\infty}\mathrm{d}x\,G_\sigma^{(0)}(x,t|x_0)=1$.

We can now build the expression for the full probability density $G(x,\sigma,t|x_0,\sigma_0)$ in terms of the switching probability density $G_S(x,\sigma,t|x_0,\sigma_0)$ by conditioning on the last switching event that has occurred:

\begin{equation}\label{eq:G}
\begin{aligned}
G(x,\sigma,t|x_0,\sigma )&=\Psi_\sigma(t)\,G_\sigma^{(0)}(x,t|x_0)+\int_0^t\mathrm{d}\tau\int_{-\infty}^{+\infty}\mathrm{d}y\,G_S(y,\sigma,\tau|x_0,\sigma )\,\Psi_\sigma(t-\tau)G_\sigma^{(0)}(x,t-\tau|y),\\[2mm]
G(x,\sigma,t|x_0,-\sigma)&=\int_0^t\mathrm{d}\tau\int_{-\infty}^{+\infty}\mathrm{d}y\,G_S(y,\sigma,\tau|x_0,-\sigma)\,\Psi_{\sigma}(t-\tau)G_{\sigma}^{(0)}(x,t-\tau|y).
\end{aligned}
\end{equation}

The equation for $G(x,\sigma,t|x_0,\sigma )$ can be understood as follows: the first contribution $\Psi_\sigma(t)G_\sigma^{(0)}(x,t|x_0)$ corresponds to trajectories with no switching events in the interval $(0,t)$, while the second to trajectories that display a last switch at position $y$ at time $\tau$ with the subsequent switch occurring after $t$. The expression for $G(x,\sigma,t|x_0,-\sigma)$ follows from the same reasoning. 

The knowledge of $G(x,\sigma,t|x_0,\sigma_0)$ allows us to calculate $\rho_\sigma(x,t|x_0)$, the probability density to be in $(x,\sigma )$ at time $t$ given the initial position $x_0$, by marginalizing with respect to the initial state, i.e., 

\begin{equation}\label{eq:rhoi}
\rho_\sigma(x,t|x_0)=\sum_{\sigma_0}\lambda_{\sigma_0} \,G(x,\sigma,t|x_0,\sigma_0),
\end{equation}
where the initial state is given by $\sigma_0=+$ with probability $\lambda\in[0,1]$, and $\sigma_0=-$ with probability $1-\lambda$, for which we adopt the compact notation $\lambda_{\sigma_0}\equiv \left[1-\sigma_0(1-2\lambda)\right]/2$. 
The total density $\rho(x,t|x_0)$ is then given by
\begin{equation}
    \rho(x,t|x_0)=\sum_\sigma \rho_\sigma(x,t|x_0).
\end{equation}

The expressions of both  $G(x,\sigma,t|x_0,\sigma_0)$ and $G_S(x,\sigma,t|x_0,\sigma_0)$ can be simplified by exploiting the properties of the Laplace transform, that we denote as $\mathcal{L}\{f(t)\}(s)\equiv \widetilde{f}(s)=\int_0^\infty \mathrm{d}t\,e^{-st}f(t)$. In particular, due to the convolution theorem of the Laplace transform, the time integrals in Eqs. \eqref{eq:GS} and \eqref{eq:GS2} become a product of the Laplace transform of $G_\sigma(x,t|x_0)$, namely

\begin{equation}\label{eq:GSLT}
\begin{aligned}
\widetilde{G}_S(x,\sigma,s|x_0,\sigma )=&\sum_{n=1}^\infty\int_{-\infty}^{+\infty}\left(\prod_{l=1}^{2n-1}\mathrm{d}z_l\right)\,\left(\prod_{k=1}^{n}\widetilde{G}_\sigma(z_{2k-1},s|z_{2k-2})\,\widetilde{G}_{-\sigma}(z_{2k},s|z_{2k-1})\right)\\
=&\int_{-\infty}^{+\infty}\mathrm{d}y\,\widetilde{G}_{-\sigma}(x,s|y)\widetilde{G}_S(y,-\sigma,s|x_0,\sigma ),\\
\widetilde{G}_S(x,-\sigma,s|x_0,\sigma )=&\,\widetilde{G}_\sigma(x,s|x_0)+\int_{-\infty}^{+\infty}\mathrm{d}y\,\widetilde{G}_\sigma(x,s|y)\widetilde{G}_S(y,\sigma,s|x_0,\sigma ),
\end{aligned}
\end{equation}
where we identify $z_0\equiv x_0$ and $z_{2n}\equiv x.$

Equation \eqref{eq:GSLT} allows us to calculate $\widetilde{P}_S(\sigma,s|\sigma_0)$, the Laplace transform of $P_S(\sigma,t|\sigma_0)$, the probability to have a switch at time $t$ given the initial state $\sigma$, by integrating $\widetilde{G}_S(x,\sigma, s|x_0,\sigma_0)$ in Eq. \eqref{eq:GSLT} over $x$:

\begin{equation}\label{eq:norms}
\begin{aligned}
\widetilde{P}_S(\sigma,s|\sigma )&\equiv\int_{-\infty}^{+\infty}\mathrm{d}x\,\widetilde{G}_S(x,\sigma,s|x_0,\sigma )=\sum_{n=1}^\infty \left[\widetilde{\psi}_+(s)\widetilde{\psi}_-(s)\right]^n=\frac{\widetilde{\psi}_+(s)\widetilde{\psi}_-(s)}{1-\widetilde{\psi}_+(s)\widetilde{\psi}_-(s)},\\
\widetilde{P}_S(-\sigma,s|\sigma )&\equiv\int_{-\infty}^{+\infty}\mathrm{d}x\,\widetilde{G}_S(x,-\sigma,s|x_0,\sigma )=\widetilde{\psi}_\sigma(s)\sum_{n=0}^\infty \left[\widetilde{\psi}_+(s)\widetilde{\psi}_-(s)\right]^n=\frac{\widetilde{\psi}_\sigma(s)}{1-\widetilde{\psi}_+(s)\widetilde{\psi}_-(s)},
\end{aligned}
\end{equation}
where the first equality follows from the integral relation
\begin{equation}
    \mathcal{L}\left\{\int_{-\infty}^{+\infty}\mathrm{d}x\,\psi_\sigma(t)\,G_\sigma^{(0)}(x,t|x_0)\right\}(s)=\widetilde{\psi}_\sigma(s),
\end{equation}
while the convergence of the geometric series is ensured by the fact that  $|\psi_\sigma(s)|<1$ for $\operatorname{Re}(s)>0$.

Similarly, the Laplace transform of Eq. \eqref{eq:G} for the full probability density $G(x,\sigma,t|x_0,\sigma_0)$ reduces to
\begin{equation}\label{eq:GLT}
\begin{aligned}
\widetilde{G}(x,\sigma,s|x_0,\sigma )&=\mathcal{L}\left\{\Psi_\sigma(t)\,G_\sigma^{(0)}(x,t|x_0)\right\}(s)+\int_{-\infty}^{+\infty}\mathrm{d}y\,\widetilde{G}_S(y,\sigma,s|x_0,\sigma )\,\mathcal{L}\left\{\Psi_\sigma(t)\,G_\sigma^{(0)}(x,t|y)\right\}(s),\\
\widetilde{G}(x,\sigma,s|x_0,-\sigma)&=\int_{-\infty}^{+\infty}\mathrm{d}y\,\widetilde{G}_S(y,\sigma,s|x_0,-\sigma)\,\mathcal{L}\left\{\Psi_{\sigma}(t)\,G_{\sigma}^{(0)}(x,t|y)\right\}(s).
\end{aligned}
\end{equation}
By integrating $\widetilde{G}(x,\sigma,s|\sigma_0,x_0)$ we derive  $\widetilde{P}(\sigma,s|\sigma_0)$, the Laplace transform of  the probability $P(\sigma,t|\sigma_0)$ of $\sigma(t)$ conditioned on the initial state $\sigma_0$:

\begin{equation}\label{eq:Psc}
\begin{aligned}
\widetilde{P}(\sigma,s|\sigma )&\equiv\int_{-\infty}^{+\infty}\mathrm{d}x\,\widetilde{G}(x,\sigma,s|x_0,\sigma )=\frac{\widetilde{\Psi}_\sigma(s)}{1-\widetilde{\psi}_+(s)\widetilde{\psi}_-(s)},\\[2mm]
\widetilde{P}(\sigma,s|-\sigma)&\equiv\int_{-\infty}^{+\infty}\mathrm{d}x\,\widetilde{G}(x,\sigma,s|x_0,-\sigma)=\frac{\widetilde{\Psi}_{\sigma}(s)\widetilde{\psi}_{-\sigma}(s)}{1-\widetilde{\psi}_+(s)\widetilde{\psi}_-(s)}.
\end{aligned}
\end{equation}
The expressions above allows us to compute the Laplace transform of the probability $P(\sigma,t)$ to find a particle in the state $\sigma$ at time $t$, that is
\begin{equation}\label{eq:Ps}
\begin{aligned}
\widetilde{P}(\sigma,s)\equiv\int_{-\infty}^{+\infty}\mathrm{d}x\,\widetilde{\rho}_\sigma(x,s|x_0)&=\frac{\widetilde{\Psi}_\sigma(s)}{1-\widetilde{\psi}_+(s)\widetilde{\psi}_-(s)}\left[\lambda_\sigma +\lambda_{-\sigma}\widetilde{\psi}_{-\sigma}(s)\right].
\end{aligned}
\end{equation}
As a simple check, we compute the overall normalization as $\int_{-\infty}^{+\infty}\mathrm{d}x\,\widetilde{\rho}(x,s|x_0)=1/s$ as it should, where we use the fact that $\widetilde{\Psi}_\sigma(s)=\left[1-\widetilde{\psi}_\sigma(s)\right]/s.$

We conclude this Section by mentioning the fact that the calculations of the various  quantities considered so far do not require the process $c(t)$ to be symmetric, but only on the fact that it is a two-state process. Indeed, one can use the expressions also for the asymmetric process $c(t)$ taking the two values $c_+$ and $c_-$, by simply identifying the sign $\sigma(t)$ with the subscript of $c_\sigma.$ Henceforth, we assume $c(t)$ to be, in general, asymmetric.

\subsection{Differential description}

We now show how to derive, by means of the integral representation in Eqs. \eqref{eq:GS}, \eqref{eq:GS2} and \eqref{eq:G}, the Fokker-Planck equation for the probability $G(x,t,\sigma|x_0,\sigma_0)$.
Between two switches, the process $x(t)$ in Eq. \eqref{eq:langevin} coincides with a Ornstein-Uhlenbeck process. Accordingly, the probability density $G_\sigma^{(0)}(x,t|x_0)$ satisfies the Fokker-Planck equation
\begin{equation}\label{eq:FPG0}
\begin{aligned}
\frac{\partial G_\sigma^{(0)}(x,t|x_0)}{\partial t}&=\hat{O}_\sigma(x)G_\sigma^{(0)}(x,t|x_0)\\
&=D\frac{\partial^2 G_\sigma^{(0)}(x,t|x_0)}{\partial x^2}+\nu\frac{\partial}{\partial x}\left[(x-c_\sigma)G_\sigma^{(0)}(x,t|x_0)\right],
\end{aligned}
\end{equation}
with the initial condition $G_\sigma^{(0)}(x,0|x_0)=\delta(x-x_0)$, where the Fokker-Planck operator $\hat{O}_\sigma(x)$ is defined by the second equality of the equation above.
By differentiating $G(x,\sigma,t|x_0,\sigma)$ in Eq. \eqref{eq:G} with respect to the final time $t$, we get
\begin{equation}\label{eq:G_time_pp}
\begin{aligned}
\frac{\partial G(x,\sigma,t|x_0,\sigma)}{\partial t}=&-G_\sigma(x,t|x_0)+\hat{O}_\sigma(x)\,\Psi_\sigma(t)G_\sigma^{(0)}(x,t|x_0)\\
&+\int_{-\infty}^{+\infty}\mathrm{d}y\,G_S(y,\sigma,t|x_0,\sigma)\Psi_\sigma(0)G_\sigma^{(0)}(x,0|y)\\
&+\int_0^t\mathrm{d}\tau\,\int_{-\infty}^{+\infty}\mathrm{d}y\,G_S(y,\sigma,\tau|x_0,\sigma)\left[-G_\sigma(x,t-\tau|y)+\Psi_\sigma(t-\tau)\hat{O}_\sigma (x) \,G_\sigma^{(0)}(x,t-\tau|y)\right]\\
=&\,\hat{O}_\sigma(x)\,G(x,\sigma,t|x_0,\sigma)-G_S(x,-\sigma,t|x_0,\sigma)+G_S(x,\sigma,t|x_0,\sigma).
\end{aligned}
\end{equation}
In the first line of the equation above we have used the relation $\partial_t \Psi_\sigma(t)=-\psi_\sigma(t)$ following from Eq. \eqref{eq:Psi}, and Eq. \eqref{eq:FPG0}.
In the second line, we substitute the normalization condition of the waiting-time distribution $\Psi_\sigma(0)=1$, the initial condition $G_\sigma^{(0)}(x,0|x_0)=\delta(x-x_0)$, and the integral expression of $G_S(x,\sigma,t|x_0,\sigma_0)$ in Eqs. \eqref{eq:GS} and \eqref{eq:GS2}.

Following the same steps as above for $G(x,\sigma,t|x_0,\sigma_0)$, it is possible to generalize Eq. \eqref{eq:G_time_pp} to any initial state $\sigma_0$:

\begin{equation}\label{eq:G_time}
\frac{\partial G(x,\sigma,t|x_0,\sigma)}{\partial t}=\hat{O}_\sigma(x)\,G(x,\sigma,t|x_0,\sigma_0)+G_S(x,\sigma,t|x_0,\sigma_0)-G_S(x,-\sigma,t|x_0,\sigma_0).
\end{equation}
As expressed in Eq. \eqref{eq:G_time}, the time evolution of $G(x,\sigma,t|x_0,\sigma_0)$ is due to two mechanisms: the first term $\hat{O}_\sigma(x)\,G(x,\sigma,t|x_0,\sigma_0)$ corresponds to the evolution of particles in $(x,\sigma)$ at $t$ according the Orstein-Uhlenbeck dynamics; the second, corresponds to the net flux of particle that switch in or out from the state $(x,\sigma)$ at $t$ given the initial configuration $(x_0,\sigma_0)$, expressed via the switching probability density  $G_S(x,\sigma,t|x_0,\sigma_0)$.
Note that the Fokker-Planck description provided in this Section is independent of the fact that $G^{(0)}_\sigma(x,t|x_0)$ represents the probability density of the Ornstein-Uhlenbeck process. Indeed, it is sufficient that $G^{(0)}_\sigma(x,t|x_0)$ obeys Eq. \eqref{eq:FPG0}, with $\hat{O}_\sigma(x)$ being the Fokker-Planck operator relative to the underlying processes, e.g, for any generic potential $V(x)$.



\subsection{Markovian limit}\label{App:Markovian}
We now focus our analysis on the statistical properties of the process $x(t)$ when it is Markovian case, i.e., when the waiting times are exponentially distributed $\psi_\sigma(\tau)=r_\sigma \,e^{-r_\sigma\tau}$. Markovianity results from the fact that the center $c(t)$ switches to $c(t+\mathrm{d}t)$ with constant a rate $r_\sigma$ in time. Accordingly, the probability density $G_S(x,\sigma,t|x_0,\sigma_0)$ is simply related to $G(x,\sigma,t|x_0,\sigma_0)$ as
\begin{equation}\label{eq:GS_exp}
    G_S(x,\sigma,t|x_0,\sigma_0)=r_{-\sigma}G(x,-\sigma,t|x_0,\sigma_0),
\end{equation}
as a consequence of the fact that switching at a given time $t$ depends only on the current state $\sigma(t)$. From a mathematical point of view, Eq.~\eqref{eq:GS_exp} is a consequence of the identity  $\psi_\sigma(\tau)=r_\sigma\Psi_\sigma(\tau)$.

By substituting Eq.~\eqref{eq:GS_exp} in Eq.~\eqref{eq:G_time}, we get closed differential equations for the probability density $G(x,\sigma,t|x_0,\sigma_0)$, namely,

\begin{equation}\label{eq:Gres}
    \frac{\partial G(x,\sigma,t|x_0,\sigma)}{\partial t}=\hat{O}(x)\,G(x,\sigma,t|x_0,\sigma_0)-r_{\sigma}G(x,\sigma,t|x_0,\sigma_0)+r_{-\sigma}G(x,-\sigma,t|x_0,\sigma_0),
\end{equation}
with the initial condition $G(x,\sigma,0|x_0,\sigma_0)=\delta_{\sigma\sigma_0}\delta(x-x_0).$
Furthermore, the probability density $\rho_\sigma(x,t|x_0)$ in Eq. \eqref{eq:rhoi} also satisfies Eq. \eqref{eq:Gres}, due to the linearity of its definition. Accordingly, the stationary distribution $\rho_\sigma^{\rm st}(x)$ satisfies the equation

\begin{equation}
    \hat{O}_\sigma(x)\rho_\sigma^{\rm st}(x)=r_\sigma \rho_\sigma^{\rm st}(x)-r_{-\sigma}\rho_{-\sigma}^{\rm st}(x),
\end{equation}
whose solution is given by Eq. \eqref{eq:rhosol}.
\subsubsection{Transition between an Unimodal and a  bimodal distribution}

\begin{figure}
\includegraphics[width = 0.5\linewidth]{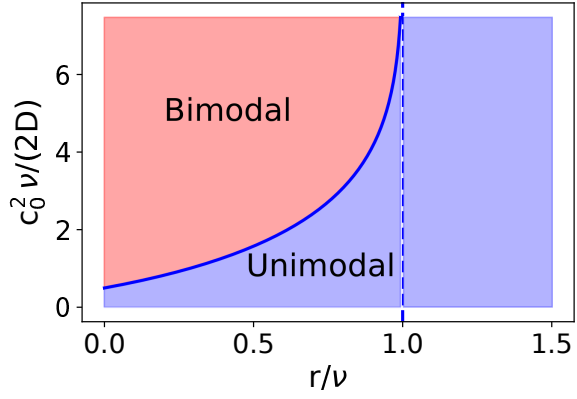}
\caption{Character of the stationary distribution $\rho^{\rm st}(x)$ depending on the values of the two parameters indicated in the plot. The blue area corresponds to an unimodal distribution, while the blue one to a bimodal regime. The two regions are delimited by the critical line $\chi^*(\zeta=r/\nu)$ in Eq. \eqref{eq:chistar} which diverges at $r=\nu$ (dashed vertical line).}\label{Fig:chiq}
\end{figure}

A possible way to ascertain whether the process $x(t)$ displays oscillatory behavior is to look at the unimodal character of its stationary density

\begin{equation}
    \rho^{\rm st}(x)=\rho^{\rm st}_+(x)+\rho^{\rm st}_-(x).
\end{equation}
In the symmetric case $r_\sigma=r$, it is possible to characterize analytically the transition from unimodal to bimodal stationary density $\rho^{\rm st}(x)$. These two regimes depend on the values of the parameters of the model, as shown in Fig. \ref{fig:rho} numerically and analytically on the basis of Eq. \eqref{eq:rhosol}. In particular, bistability emerges whenever the relaxation is fast enough with respect to the switching frequency, i.e., $\tau_\nu\ll \langle \tau\rangle_\pm$. 
The regime of the system is identified by studying whether the origin $x=0$ is a point of local maximum or minimum for $\rho^{\rm st}(x)$.
In the former case, $\rho^{\rm st}(0)$ displays an unique global maximum and it is unimodal, see the red and blue curves in Fig.~\ref{fig:rho}.
In the latter, $\rho^{\rm st}(0)$ is bimodal, it shows a local minimum in the origin, and two symmetric maxima, see the green curve of Fig. \ref{fig:rho}.
The symmetric solution $\rho^{\rm st}(x)$ to Eq. \eqref{eq:rhosol} is given by:

\begin{equation}\label{eq:rostsym}
\rho^{\rm st}(x)=\frac{1}{\sqrt{\pi}}\frac{\Gamma\left(\zeta+\frac{1}{2}\right)}{\Gamma\left(\zeta-\frac{1}{2}\right)}\int_{-1}^{+1}\mathrm{d}z \,\rho_{\rm G}(x-c_0z) (1-z^2)^{\zeta-1},
\end{equation}
where $\Gamma$ denotes the Gamma function, $\zeta=r/\nu$, $\rho_G(x)$ is defined to express the asymmetric solution in Eq. \eqref{eq:rhosol}, and we set $c_\sigma=\sigma c_0$, for simplicity.
First, we calculate the derivative of the stationary distribution  $\rho^{\rm st}(x)$  
\begin{equation}
\frac{\partial \rho^{\rm st}(x)}{\partial x}=\frac{\nu}{D\sqrt{\pi}}\frac{\Gamma\left(\zeta+\frac{1}{2}\right)}{\Gamma\left(\zeta-\frac{1}{2}\right)}\int_{-1}^{+1}\mathrm{d}z \,\rho_{\rm G}(x-c_0z) (1-z^2)^{\zeta-1}(c_0z-x),
\end{equation}
which, due to the integrand being an odd function of $z$, vanishes at $x=0$ as expected, confirming that this point is always a point of maximum or minimum for $\rho^{\rm st}(x)$. In order to understand its actual nature, we study the sign of the second derivative of $\rho^{\rm st}(x)$ at $x=0$:


\begin{equation}
\frac{\partial^2 \rho^{\rm st}(x)}{\partial x^2}\Bigg|_{x=0}=\frac{1}{\sqrt{2\pi(D/\nu)^3}}\left[\frac{\chi}{\zeta+1/2}\,\,{}_1F_1\left(\frac{3}{2},\zeta+\frac{3}{2},-\chi\right)-{}_1F_1\left(\frac{1}{2},\zeta+\frac{1}{2},-\chi\right)\right],
\end{equation}
where we recall $\chi=c_0^2\nu/(2D)$, and ${}_1F_1$ denotes the confluent hypergeometric function. Accordingly, the transition occurs upon crossing the critical value $\chi^*$ found by imposing $\partial_x^2\rho^{\rm st}(0)=0$, i.e.,
\begin{equation}\label{eq:chistar}
    \chi^*=\frac{(\zeta+1/2)\,\,{}_1F_1\left(1/2,\zeta+1/2,-\chi^*\right)}{{}_1F_1\left(3/2,\zeta+3/2,-\chi^*\right)}.
\end{equation}
In general, if $\chi<\chi^*$ then $\partial_x^2\rho^{\rm st}(0)$ is negative and $\rho^{\rm st}(x)$ is unimodal, while it is bimodal otherwise.
More specifically, for $r\geq \nu$ ($\zeta\geq 1$), the value of $\chi^*$ diverges and  $\rho^{\rm st}(x)$ is always unimodal. For $r<\nu$ ($\zeta<1$), the critical $\chi^*$ is finite and grows monotonically upon increasing $\zeta$, as shown in Fig. \ref{Fig:chiq}.

\subsection{Moments of $G_S(x,\sigma,t|x_0,\sigma_0)$}

For later convenience, we now derive the Laplace transform of the first and second moment of the switching probability density $G_S(x,\sigma,t|x_0,\sigma_0)$. We start by calculating the expectation value of the probability density $G_\sigma(x,t|x_0)=\psi_\sigma(t)G_\sigma^{(0)}(x,t|x_0)$. As reported above, $G_\sigma^{(0)}(x,t|x_0)$ is the probability density of the Ornstein-Uhlenbeck process in Eq. \eqref{eq:langevin}
with fixed $c(t)=c_\sigma$, which is given by a Gaussian with average
\begin{equation}\label{eq:musigma}
\mu_\sigma(t|x_0)\equiv c_\sigma(1-e^{-\nu t})+x_0\,e^{-\nu t},
\end{equation}
and variance 
\begin{equation}\label{eq:Sigmasigma}
\Sigma^2(t)\equiv \frac{D}{\nu}(1-e^{-2\nu t}).
\end{equation}
Given the time convolution structure of Eqs. \eqref{eq:GS} and \eqref{eq:GS2}, it is natural to continue the calculations of the moments of $G_S(x,\sigma,t|x_0,\sigma_0)$ in the Laplace transform. In particular, for the Laplace transform of the first moment of $G_\sigma(x,t|x_0)$ we find
\begin{equation}\label{eq:xpmLT}
\begin{aligned}
\langle \widetilde{x}_\sigma(s|x_0)\rangle\equiv&\int_{-\infty}^{+\infty}\mathrm{d}z\,z\,\widetilde{G}_\sigma(z,s|x_0)=\mathcal{L}\left\{\psi_\sigma(t)\mu_\sigma(t|x_0)\right\}\\
=&\,\mathcal{L}\left\{\psi_\sigma(t)\left[x_0 e^{-\nu t}+ c_\sigma (1-e^{-\nu t})\right]\right\}\\
=&\,x_0\,\widetilde{\psi}_\sigma(s+\nu)+ c_\sigma\left[\widetilde{\psi}_\sigma(s)-\widetilde{\psi}_\sigma(s+\nu)\right]\\
=&\,x_0\,a_\sigma^{(1)} + \,b_\sigma^{(1)},
\end{aligned}
\end{equation}
where we define the auxiliary quantities $a_\sigma^{(n)}\equiv\widetilde{\psi}_\sigma(s+n\nu) $ and $b_\sigma^{(1)}\equiv  c_\sigma\left[\widetilde{\psi}_\sigma(s)-\widetilde{\psi}_\sigma(s+\nu)\right]$. The second line of Eq. \eqref{eq:xpmLT} is found by direct substitution of $\mu_\sigma(t|x_0)$ in Eq. \eqref{eq:musigma}, while the third by applying the property of the Laplace transform  $\mathcal{L}\left\{e^{-\nu t}f(t)\right\}(s)=\widetilde{f}(s+\nu)$.

\subsubsection{First moment}

We are now in the position to calculate $\langle\widetilde{x}(\sigma,s|x_0,\sigma)\rangle$, the Laplace transform of the first moment of $G_S(x,t,\sigma|x_0,\sigma)$ in Eq. \eqref{eq:GSLT}:
\begin{equation}\label{eq:xgspp}
\begin{aligned}
\langle\widetilde{x}_S(\sigma,s|x_0,\sigma)\rangle=&\int_{-\infty}^{+\infty}\mathrm{d}y\,y\,\widetilde{G}_S(y,\sigma,s|x,\sigma)=\sum_{n=1}^\infty I^{(1)}_\sigma (n,s)\\     =&\sum_{n=1}^\infty\int_{-\infty}^{+\infty}\left(\prod_{k=1}^{2n-2}\mathrm{d}z_k\right)\,\left[\prod_{k=1}^{n-1}\widetilde{G}_\sigma(z_{2k-1},s|z_{2k-2})\,\widetilde{G}_{-\sigma}(z_{2k},s|z_{2k-1})\right]\\
&\times \int_{-\infty}^{+\infty}\mathrm{d}z_{2n-1}\,\widetilde{G}_\sigma(z_{2n-1},s|z_{2n-2})\,\langle\widetilde{x}_{-\sigma}(s|z_{2n-1})\rangle\\
=&\sum_{n=1}^\infty\int_{-\infty}^{+\infty}\left(\prod_{k=1}^{2n-2}\mathrm{d}z_k\right)\,\left[\prod_{k=1}^{n-1}\widetilde{G}_\sigma(z_{2k-1},s|z_{2k-2})\,\widetilde{G}_{-\sigma}(z_{2k},s|z_{2k-1})\right]\left(z_{2n-1}A^{(1)}+B^{(1)}_\sigma\right)\\
=&\sum_{n=1}^\infty\left[I_\sigma ^{(1)}(n-1,s) A^{(1)}+B^{(1)}_\sigma (A^{(0)})^{n-1}\right],\\
\end{aligned}
\end{equation}
where we define $A^{(n)}\equiv a_+^{(n)}a_-^{(n)}$, and $B^{(1)}_\sigma\equiv b_\sigma^{(1)}a_{-\sigma}^{(1)}+b_{-\sigma}^{(1)}\widetilde{\psi}_\sigma(s) $. The definition of the integral $I_\sigma^{(1)}(n,s)$ is given by the second and third line of Eq. \eqref{eq:xgspp}, and results from the insertion of the expression of $G_S(y,\sigma,t|x,\sigma)$ in Eq. \eqref{eq:GS}. The integral in the third line of Eq. \eqref{eq:xgspp}, evaluated via Eq. \eqref{eq:xpmLT}, coincides with $I_\sigma^{(1)}(1,s)=x_0A^{(1)}+B_\sigma^{(1)}$ in the case where the initial position coincides with the integration variable $z_{2n-1}$. In the last line we recognize the same type of integral $I_\sigma^{(1)}(n-1,s)$ as in the second and third line, plus the extra term $(A^{(0)})^{n-1}=[\widetilde{\psi}_\sigma(s)\widetilde{\psi}_{-\sigma}(s) ]^{n-1}$
which follows from $\int_{-\infty}^{+\infty}\mathrm{d}x\,\widetilde{G}_\sigma(x,s|x_0)=\widetilde{\psi}_\sigma(s)$.

The integral $I_\sigma^{(1)}(n,s)$ can be evaluated by recursively substituting its lower $n$-degree expression, down to the known quantity  $I_\sigma^{(1)}(1,s)$:

\begin{equation}\label{eq:I1ns}
\begin{aligned}
I^{(1)}_\sigma (n,s)&=I_\sigma ^{(1)}(n-1,s) A^{(1)}+B^{(1)}_\sigma (A^{(0)})^{n-1}\\
&=\left(A^{(1)}\right)^k\,I_\sigma ^{(1)}(n-k,s)+B^{(1)}_\sigma\sum_{l=0}^{k-1}\left(A^{(1)}\right)^l\left(A^{(0)}\right)^{n-1-l}\\
&=x_0\,\left(A^{(1)}\right)^n+B^{(1)}_\sigma\sum_{l=0}^{n-1}\left(A^{(1)}\right)^l\left(A^{(0)}\right)^{n-1-l}\\
&=x_0\,\left(A^{(1)}\right)^n+B^{(1)}_\sigma\frac{\left(A^{(0)}\right)^{n}-\left(A^{(1)}\right)^{n}}{A^{(0)}-A^{(1)}}.
\end{aligned}
\end{equation}
Finally, we substitute the expression \eqref{eq:I1ns} of $I_\sigma^{(1)}(n,s)$  into Eq. \eqref{eq:xgspp}, determining the Laplace transform of the first moment of the switching probability density $G_S(x,\sigma,t|x_0,\sigma_0)$:
\begin{equation}\label{eq:xspp}
\begin{aligned}
\langle\widetilde{x}_S(\sigma,s|x_0,\sigma)\rangle &=\sum_{n=1}^{\infty}I_\sigma^{(1)}(n,s)\\
&=\sum_{n=1}^{\infty}\left[x_0\,\left(A^{(1)}\right)^n+B^{(1)}_\sigma\frac{\left(A^{(0)}\right)^{n}-\left(A^{(1)}\right)^{n}}{A^{(0)}-A^{(1)}}\right]\\
&=x_0\frac{A^{(1)}}{1-A^{(1)}}+\frac{B^{(1)}_\sigma}{(1-A^{(0)})(1-A^{(1)})},
\end{aligned}
\end{equation}
where last equality follows by summing the geometric series, whose convergence is ensured by $|A^{(n)}|<1$ for $\operatorname{Re}(s)>0$. In general, it is easy to check that $A^{(n)}$ satisfies the inequality $|A^{(n)}|<|A^{(m)}|$ for $m<n$.

Analogously to what was done for $\langle\widetilde{x}_S(\sigma,s|x_0,\sigma)\rangle$, 
we can compute $\langle\widetilde{x}_S(-\sigma,s|x_0,\sigma)\rangle$, 
the first moment of $\widetilde{G}_S(x,-\sigma,s|x_0,\sigma)$ in Eq. \eqref{eq:GSLT}, as
\begin{equation}\label{eq:xspm}
\begin{aligned}
\langle\widetilde{x}_S(-\sigma,s|x_0,\sigma)\rangle&=\int_{-\infty}^{+\infty}\mathrm{d}y\,y\,\widetilde{G}_S(y,-\sigma,s|x_0,\sigma)\\
&=\langle\widetilde{x}_\sigma(s|x_0)\rangle+\int_{-\infty}^{+\infty}\mathrm{d}y\,\langle\widetilde{x}_\sigma(s|y)\rangle\,\widetilde{G}_S(y,\sigma,s|x_0,\sigma)\\
&=x_0\,a_\sigma^{(1)} + \,b_\sigma^{(1)}+\int_{-\infty}^{+\infty}\mathrm{d}y\,\left(y\,a_\sigma^{(1)} +\,b_\sigma^{(1)}\right)\widetilde{G}_S(y,\sigma,s|x_0,\sigma)\\
&=x_0\,a_\sigma^{(1)} + \,b_\sigma^{(1)}+a_\sigma^{(1)}\left[x_0\frac{A^{(1)}}{1-A^{(1)}}+\frac{B^{(1)}_\sigma}{(1-A^{(0)})(1-A^{(1)})}\right]+b_\sigma^{(1)}\frac{A^{(0)}}{1-A^{(0)}}\\
&=\frac{a_\sigma^{(1)}}{1-A^{(1)}}\left[x_0+\frac{B^{(1)}_\sigma}{1-A^{(0)}}\right]+\frac{b_\sigma^{(1)}}{1-A^{(0)}}.
\end{aligned}
\end{equation}

For completeness, we evaluate the stationary value of $\langle\widetilde{x}_S(-\sigma,s|x_0,\sigma)\rangle$ 
and $\langle\widetilde{x}_S(\sigma,s|x_0,\sigma)\rangle$ 
by using the final value theorem of the Laplace transform, i.e.,

\begin{equation}
\begin{aligned}
\langle x_S \rangle_\sigma^{\rm st}&\equiv\lim_{t\rightarrow \infty}\langle x_S(\sigma,t|\sigma_0,x_0)\rangle=\lim_{s\rightarrow 0}s\, \langle\widetilde{x}_S(\sigma,s|\sigma_0,x_0)\rangle\\
&=\frac{c_\sigma\left[1-\widetilde{\psi}_\sigma(\nu)\right]\widetilde{\psi}_{-\sigma}(\nu)+c_{-\sigma}\left[1-\widetilde{\psi}_{-\sigma}(\nu)\right]}{2\langle \tau\rangle\,\left[1-\widetilde{\psi}_+(\nu)\widetilde{\psi}_-(\nu)\right]}.
\end{aligned}
\end{equation}
Note that the value $\langle\widetilde{x}_S(\sigma,s|\sigma_0,x_0)\rangle$ depends only on the final state, on the Laplace transform of the waiting-time distribution $\widetilde{\psi}_\sigma(\nu)$ computed at $\nu$, and on its average period 

\begin{equation}\label{eq:avtau}
    \langle \tau\rangle=\frac{\langle \tau\rangle_++\langle \tau\rangle_-}{2}.
\end{equation}

\subsubsection{Second moment}

The Laplace transform $\langle\widetilde{x}_S^2(\sigma,s|x_0,\sigma)\rangle$ of the second moments  
of $G_S(x,\sigma,t|x_0,\sigma_0)$ are computed following the same steps as those we followed above for  $\langle\widetilde{x}_S(\sigma,s|\sigma_0,x_0)\rangle$. First, we consider the second moment of the probability density $\widetilde{G}_\sigma(x,s|x_0)$, which reads
\begin{equation}\label{eq:x2pmLT}
\begin{aligned}
\langle \widetilde{x}^2_\sigma(s|x_0)\rangle\equiv&\int_{-\infty}^{+\infty}\mathrm{d}z\,z^2\,\widetilde{G}_\sigma(z,s|x_0)=\mathcal{L}\left\{\psi_\sigma(t)\left[\Sigma^2(t)+\mu^2_\sigma(t|x_0)\right]\right\}\\
=&\,\mathcal{L}\left\{\psi_\sigma(t)\left[x_0^2 e^{-2\nu t}+ c_\sigma^2 (1-2e^{-\nu t}+e^{-2\nu t})+2x_0\,c_\sigma\left(e^{-\nu t}-e^{-2\nu t}\right)+\frac{D}{  \nu}(1-e^{-2\nu t})\right]\right\}\\
=&\,x_0^2\,\widetilde{\psi}_\sigma(s+2\nu)+2x_0\,c_\sigma\left[\widetilde{\psi}_\sigma(s+\nu)-\widetilde{\psi}_\sigma(s+2\nu)\right]\\
&\,+c_\sigma^2\left[\widetilde{\psi}_\sigma(s)-2\widetilde{\psi}_\sigma(s+\nu)+\widetilde{\psi}_\sigma(s+2\nu)\right]+\frac{D}{\nu}\left[\widetilde{\psi}_\sigma(s)-\widetilde{\psi}_\sigma(s+2\nu)\right]\\
=&\,x_0^2\,a_\sigma^{(2)} + x_0\,b_\sigma^{(2)}+c^{(2)}_\sigma,
\end{aligned}
\end{equation}
where we define the auxiliary variables
\begin{equation}
\begin{aligned}
b_\sigma^{(2)}&=2\,c_\sigma\left[\widetilde{\psi}_\sigma(s+\nu)-\widetilde{\psi}_\sigma(s+2\nu)\right],\\
c_\sigma^{(2)}&=c_\sigma^2\left[\widetilde{\psi}_\sigma(s)-2\widetilde{\psi}_\sigma(s+\nu)+\widetilde{\psi}_\sigma(s+2\nu)\right]+\frac{D}{\nu}\left[\widetilde{\psi}_\sigma(s)-\widetilde{\psi}_\sigma(s+2\nu)\right].
\end{aligned}
\end{equation}
In the second line of Eq. \eqref{eq:x2pmLT} we make explicit the expression of the second moment of the position of a Ornstein-Uhlenbeck process, while, in the third we evaluate its Laplace transform.
We are now in the position to calculate $\langle\widetilde{x}_S^2(\sigma,s|x_0,\sigma)\rangle$:

\begin{equation}\label{eq:x2gspp}
\begin{aligned}
\langle\widetilde{x}_S^2(\sigma,s|x_0,\sigma)\rangle=&\int_{-\infty}^{+\infty}\mathrm{d}y\,y^2\,\widetilde{G}_S(y,\sigma,s|x,\sigma)=\sum_{n=1}^\infty I^{(2)}_\sigma (n,s)\\     
=&\sum_{n=1}^\infty\int_{-\infty}^{+\infty}\left(\prod_{k=1}^{2n-2}\mathrm{d}z_k\right)\,\left[\prod_{k=1}^{n-1}\widetilde{G}_\sigma(z_{2k-1},s|z_{2k-2})\,\widetilde{G}_{-\sigma}(z_{2k},s|z_{2k-1})\right]\\
&\times \int_{-\infty}^{+\infty}\mathrm{d}z_{2n-1}\,\widetilde{G}_\sigma(z_{2n-1},s|z_{2n-2})\,\langle\widetilde{x}^2_{-\sigma}(s|z_{2n-1})\rangle\\
=&\sum_{n=1}^\infty\int_{-\infty}^{+\infty}\left(\prod_{k=1}^{2n-2}\mathrm{d}z_k\right)\,\left[\prod_{k=1}^{n-1}\widetilde{G}_\sigma(z_{2k-1},s|z_{2k-2})\,\widetilde{G}_{-\sigma}(z_{2k},s|z_{2k-1})\right]\left(z_{2n-1}^2A^{(2)}+z_{2n-1}B^{(2)}_\sigma+C^{(2)}_\sigma\right)\\
=&\sum_{n=1}^\infty\left[I_\sigma ^{(2)}(n-1,s) A^{(2)}+I_\sigma ^{(1)}(n-1,s) B^{(2)}_\sigma+C^{(2)}_\sigma (A^{(0)})^{n-1}\right],\\
\end{aligned}
\end{equation}
where the expression of $I_\sigma^{(2)}(n,s)$ are given by the  second and third line, and we have introduced the auxiliary variables
\begin{equation}
\begin{aligned}
B^{(2)}_\sigma&\equiv b^{(2)}_\sigma a^{(2)}_{-\sigma}+b^{(2)}_{-\sigma} a^{(1)}_\sigma,\\
C^{(2)}_\sigma&\equiv c^{(2)}_\sigma a^{(2)}_{-\sigma}+a^{(0)}_\sigma c^{(2)}_{-\sigma}+b^{(2)}_{-\sigma} b^{(1)}_\sigma.
\end{aligned}
\end{equation}
In the forth line we substitute the expression  $I^{(2)}_\sigma (1,s)=x_0^2 A^{(2)}+x_0 B_\sigma^{(2)}+C_\sigma^{(2)}$, where the initial point coincides with the integration variable $z_{2n-1}$.
In the last line, we recognize the appearance of integrals of the type $I^{(2)}_\sigma (n,s)$, and $I^{(1)}_\sigma (n,s)$ at lower order in $n$.
Then, we evaluate the integral $I^{(2)}_\sigma (n,s)$ recursively as
\begin{equation}
\begin{aligned}
I^{(2)}_\sigma(n,s)\equiv& \int_{-\infty}^{+\infty}\left(\prod_{l=1}^{2n}\mathrm{d}z_l\right)\,\left[\prod_{k=1}^{n}\widetilde{G}_\sigma(z_{2k-1},s|z_{2k-2})\,\widetilde{G}_{-\sigma}(z_{2k},s|z_{2k-1})\right]\,z_{2n}^2\\
=&A^{(2)}I^{(2)}_\sigma(n-1,s)+B^{(2)}I^{(1)}_\sigma(n-1,s)+C^{(2)}_\sigma\left(A^{(0)}\right)^{n-1}\\
=&\left(A^{(2)}\right)^n x_0^2+B^{(2)}_\sigma \left(A^{(2)}\right)^{n-1}x_0+B^{(2)}_\sigma\sum_{j=0}^{n-2}\left(A^{(2)}\right)^{j} I_\sigma ^{(1)}(n-1-j,s)+C^{(2)}_\sigma\sum_{j=0}^{n-1}\left(A^{(2)}\right)^{j} \left(A^{(0)}\right)^{n-1-j}\\
=&\left(A^{(2)}\right)^n x_0^2+B^{(2)}_\sigma \left(A^{(2)}\right)^{n-1}x_0+C^{(2)}_\sigma\frac{\left(A^{(0)}\right)^{n}-\left(A^{(2)}\right)^{n}}{A^{(0)}-A^{(2)}}+x_0 B^{(0)}_\sigma A^{(1)}\frac{\left(A^{(1)}\right)^{n-1}-\left(A^{(2)}\right)^{n-1}}{A^{(1)}-A^{(2)}}\\
&+\frac{B^{(0)}_\sigma B^{(1)}_\sigma}{A^{(0)}-A^{(1)}}\left[A^{(0)}\frac{\left(A^{(0)}\right)^{n-1}-\left(A^{(2)}\right)^{n-1}}{A^{(0)}-A^{(2)}}-A^{(1)}\frac{\left(A^{(1)}\right)^{n-1}-\left(A^{(2)}\right)^{n-1}}{A^{(1)}-A^{(2)}}\right],
\end{aligned}
\end{equation}
which follows by summing the geometric sequence in the third line. Finally, by substituting the integral $I^{(2)}_\sigma(n,s)$ and $I^{(1)}_\sigma(n,s)$ in Eq. \eqref{eq:x2gspp}, we get the second moment  $\langle\widetilde{x}_S^2(\sigma,s|x_0,\sigma)\rangle$:
\begin{equation}\label{eq:xs2pp}
\begin{aligned}
\langle\widetilde{x}_S^2(\sigma,s|x_0,\sigma)\rangle=\sum_{n=1}^\infty I^{(2)}_\sigma(n,s)=&\, x_0^2\frac{A^{(2)}}{1-A^{(2)}}+x_0\frac{B^{(2)}_\sigma}{\left(1-A^{(1)}\right)\left(1-A^{(2)}\right)}\\
&+\frac{C^{(2)}_\sigma}{\left(1-A^{(0)}\right)\left(1-A^{(2)}\right)}+\frac{B^{(2)}_\sigma B^{(1)}_\sigma}{\left(1-A^{(0)}\right)\left(1-A^{(1)}\right)\left(1-A^{(2)}\right)}.
\end{aligned}
\end{equation}
The same considerations are made for $\langle\widetilde{x}_S^2(-\sigma,s|x_0,\sigma)\rangle$, 
whose calculation follows directly from the expression of $\widetilde{G}_S(x,-\sigma,s|x_0,\sigma)$ in Eq. \eqref{eq:GSLT}:

\begin{equation}\label{eq:xs2pm}
\begin{aligned}
\langle\widetilde{x}_S^2(-\sigma,s|x_0,\sigma)\rangle=&\int_{-\infty}^{+\infty}\mathrm{d}y\,y^2\,\widetilde{G}_S(y,-\sigma,s|x_0,\sigma)\\
=&\,\langle\widetilde{x}^2_\sigma(s|x_0)\rangle+\int_{-\infty}^{+\infty}\mathrm{d}y\,\langle\widetilde{x}^2_\sigma(s|y)\rangle\,\widetilde{G}_S(y,\sigma,s|x_0,\sigma)\\
=&\,x_0^2\,a_\sigma^{(2)} + x_0\,b_\sigma^{(2)}+c^{(2)}_\sigma+\int_{-\infty}^{+\infty}\mathrm{d}y\,\left(y^2\,a_\sigma^{(2)} + y\,b_\sigma^{(2)}+c^{(2)}_\sigma\right)\widetilde{G}_S(y,\sigma,s|x_0,\sigma)\\
=&\,x_0^2\,a_\sigma^{(2)} + x_0\,b_\sigma^{(2)}+\langle\widetilde{x}^2_S(\sigma,s|x_0,\sigma)\rangle \,a_\sigma^{(2)} + \langle\widetilde{x}_S(\sigma,s|x_0,\sigma)\rangle \,b_\sigma^{(2)}+\frac{c^{(2)}_\sigma}{1-A^{(0)}}\\
=&\,\frac{a_\sigma^{(2)}}{1-A^{(2)}}\left[x_0^2+x_0\frac{B^{(2)}_\sigma}{1-A^{(1)}}+\frac{C^{(2)}_\sigma}{1-A^{(0)}}+\frac{B^{(2)}_\sigma B^{(1)}_\sigma}{\left(1-A^{(0)}\right)\left(1-A^{(1)}\right)}\right]\\
&+\frac{c^{(2)}_\sigma}{1-A^{(0)}}+\frac{b_\sigma^{(2)}}{1-A^{(1)}}\left[x_0+\frac{B^{(1)}_\sigma}{1-A^{(0)}}\right].
\end{aligned}
\end{equation}
In the fifth line we substitute the expression of $\langle\widetilde{x}_S^2(\sigma,s|x_0,\sigma)\rangle$ 
in Eq. \eqref{eq:xs2pp} and of  $\langle\widetilde{x}_S(\sigma,s|x_0,\sigma)\rangle$ 
in Eq. \eqref{eq:xspp}.

For later convenience, we conclude this Section by computing the stationary limit $\langle x^2 _S\rangle_\sigma^{\rm st}$ 
of $\langle\widetilde{x}_S^2(\sigma,s|x_0,\sigma_0)\rangle$. By applying the final value theorem of the Laplace transform to Eqs. \eqref{eq:xs2pp} and \eqref{eq:xs2pm}, we get

\begin{equation}
\begin{aligned}
\langle x^2 _S\rangle_\sigma^{\rm st}&=\frac{1}{2\langle \tau\rangle\left[1-\widetilde{\psi}_+(2\nu)\widetilde{\psi}_-(2\nu)\right]}\left[C_\sigma^{(2)}(0)+\frac{B_\sigma^{(1)}(0)B_\sigma^{(2)}(0)}{1-\widetilde{\psi}_+(\nu)\widetilde{\psi}_-(\nu)}\right].
\end{aligned}
\end{equation}
In particular, we can give an explicit simple expression in the symmetric case $\psi=\psi_\sigma$, that is

\begin{equation}
\begin{aligned}
\langle x^2 _S\rangle_\sigma^{\rm st}&=\frac{1}{2\langle \tau\rangle}\left\{\frac{D}{\nu}+\frac{c_0^2}{1-\widetilde{\psi}(2\nu)}\left[1-2\widetilde{\psi}(\nu)+\widetilde{\psi}(2\nu)-2\left(\widetilde{\psi}(\nu)-\widetilde{\psi}(2\nu)\right)\frac{1-\widetilde{\psi}(\nu)}{1+\widetilde{\psi}(\nu)}\right]\right\},
\end{aligned}
\end{equation}
where $c_0\equiv(c_\sigma-c_{-\sigma})/2$ with $c_\sigma>c_{-\sigma}$.

\subsection{Moments of $G(x,\sigma,t|x_0,\sigma_0)$}
We now use the expressions of the first and second moments of the switching probability density $G_S(x,\sigma,t|x_0,\sigma_0)$ found in the previous Section to compute those of $G(x,\sigma,t|x_0,\sigma_0)$.

\begin{figure}
\centering
\includegraphics[width = 0.48\linewidth]{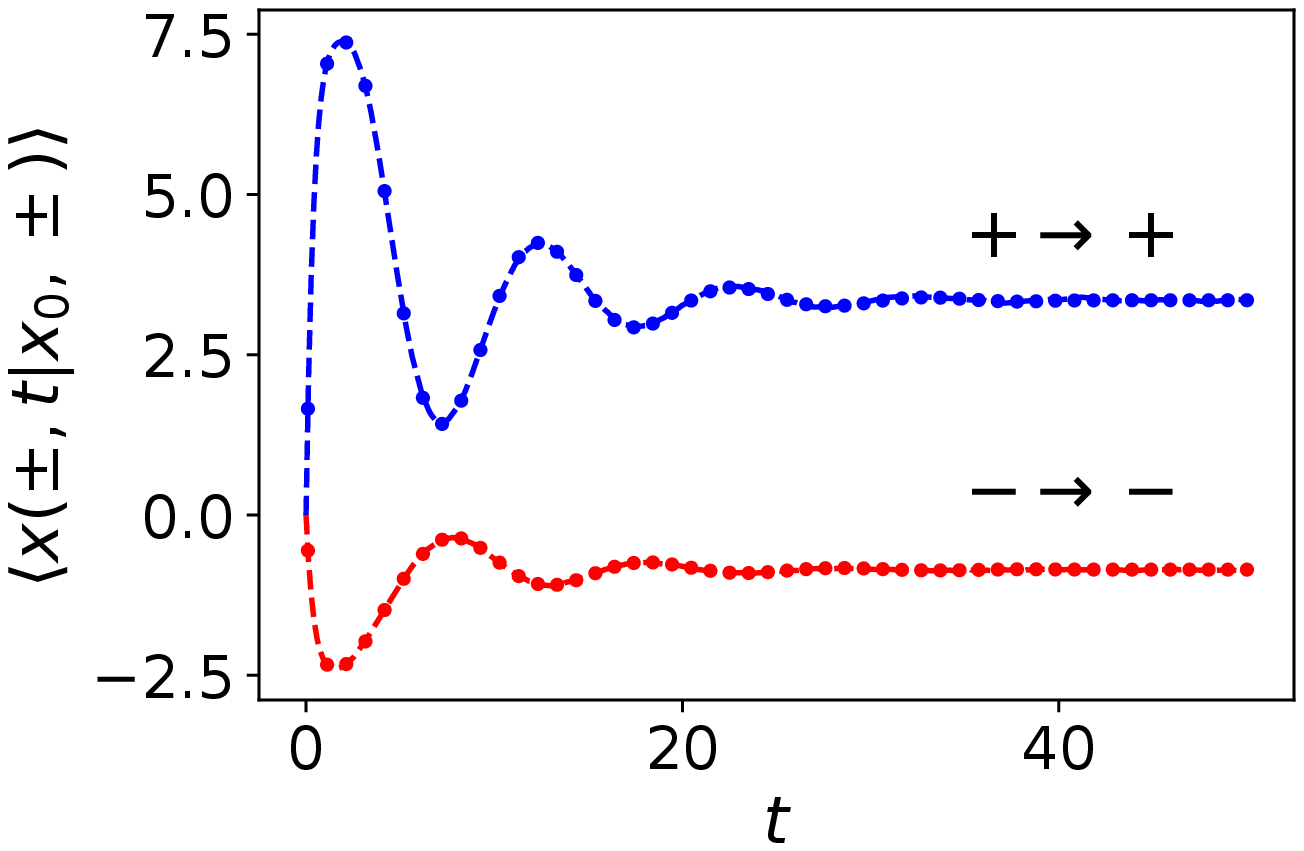}\quad \includegraphics[width = 0.48\linewidth]{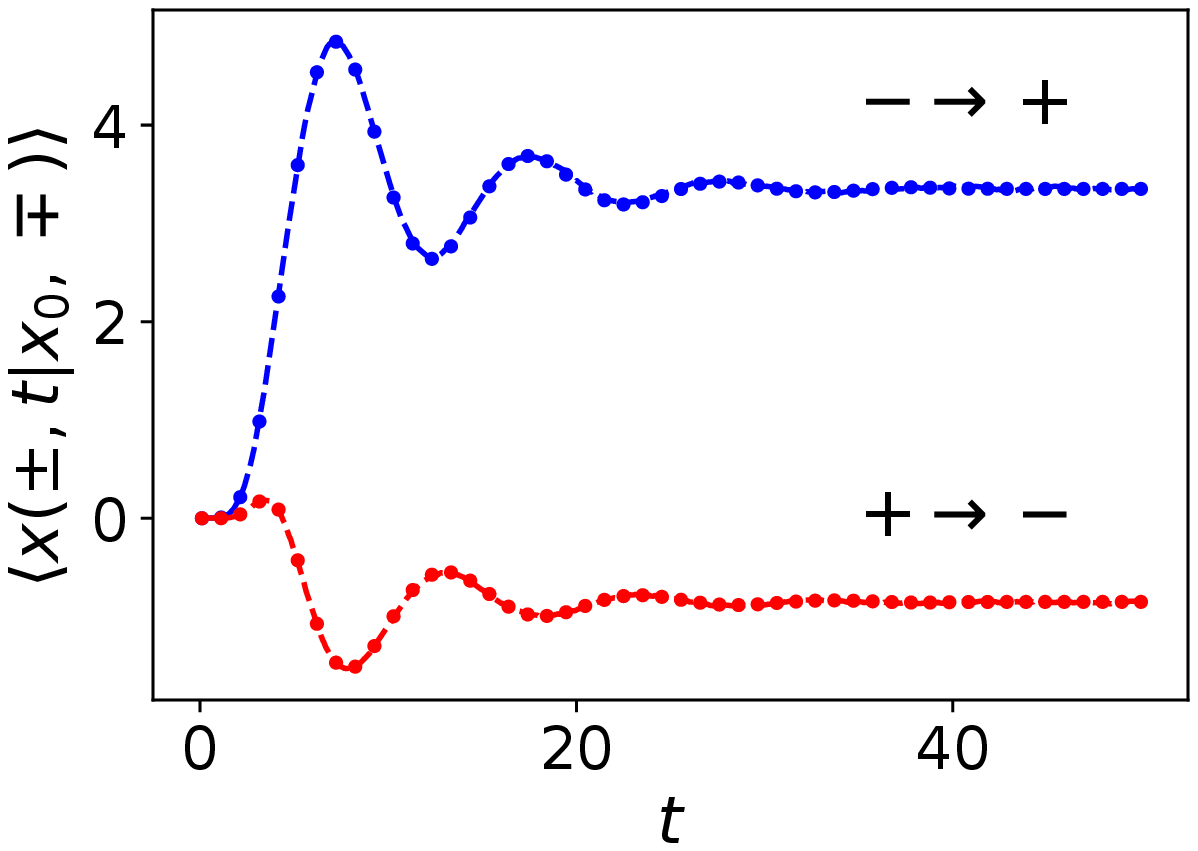}
\caption{Dependence of the conditional moments $\langle x(\sigma,t|x_0,\sigma_0)\rangle$ on time $t$ for the same (left) or different (right) initial and final potentials. In particular, in both panels the blue lines refer to a fixed potential centered in $c_+$, and the red ones to $c_-$, while  the dashed lines correspond to simulations ($N=10^5$ samples with $\Delta t=0.005$) and the dots to the  inverse Laplace transform of Eq. \eqref{eq:xpp}. Due to the initial conditions, all curves start from $x_0=0$ at time $t=0$ but, after an oscillatory transient they reach the stationary values given by Eqs. \eqref{eq:xst}. The parameters of the model are: $D=1$, $c_+=7.5$, $c_-=-2.5$, $\nu=2.5$, $k_+=10$, $\theta_+=0.5$, $k_-=5$, $\theta_-=1$, $x_0=0$ and $\lambda=0.5$.}\label{fig:xpp}
\end{figure}

\subsubsection{First moment}

As a first case, we consider the moment $\langle\widetilde{x}(\sigma,s|x_0,\sigma_0)\rangle=\int_{-\infty}^{+\infty}\mathrm{d}y\,y\,\widetilde{G}(y,\sigma,s|x_0,\sigma_0)$, that we explicitly compute by substituting into Eq. \eqref{eq:GLT} the definition of $\langle\widetilde{x}(\sigma,s|x_0,\sigma_0)\rangle$. These moments are expressed in terms of the moments of the switching probability density $G_S(x,t|x_0)$ as

\begin{equation}\label{eq:xpp}
\begin{aligned}
\langle\widetilde{x}(\sigma,s|x_0,\sigma)\rangle&=\mathcal{L}\left\{\Psi_\sigma(t)\mu_\sigma(t|x_0)\right\}(s)+\int_{-\infty}^{+\infty}\mathrm{d}y\,\widetilde{G}_S(y,\sigma,s|x_0,\sigma)\,\mathcal{L}\left\{\Psi_\sigma(t)\mu_\sigma(t|y)\right\}(s)\\
&=\widetilde{\Psi}_\sigma(s+\nu)\,\langle\widetilde{x}_S(\sigma,s|x_0,\sigma)\rangle+ c_\sigma\frac{\widetilde{\Psi}_\sigma(s)-\widetilde{\Psi}_\sigma(s+\nu)}{1-\widetilde{\psi}_+(s)\widetilde{\psi}_-(s)}+x_0\widetilde{\Psi}_\sigma(s+\nu),\\
\langle\widetilde{x}(\sigma,s|x_0,-\sigma)\rangle&=\int_{-\infty}^{+\infty}\mathrm{d}y\,\widetilde{G}_S(y,\sigma,s|x_0,-\sigma)\,\mathcal{L}\left\{\Psi_\sigma(t)\mu_\sigma(t|y)\right\}(s)\\
&=\widetilde{\Psi}_\sigma(s+\nu)\,\langle\widetilde{x}_S(\sigma,s|x_0,-\sigma)\rangle+ c_\sigma\widetilde{\psi}_{-\sigma}(s)\frac{\widetilde{\Psi}_\sigma(s)-\widetilde{\Psi}_\sigma(s+\nu)}{1-\widetilde{\psi}_+(s)\widetilde{\psi}_-(s)};
\end{aligned}
\end{equation}
a representative plot of these moments is reported in Fig. \ref{fig:xpp}.
Finally, one can reconstruct the average particle position by conditioning on the initial state as $\langle\widetilde{x}(s|x_0)\rangle=\sum_\sigma \langle\widetilde{x}(\sigma,s|x_0)\rangle $, where $\langle \widetilde{x}(\sigma,s|x_0)\rangle =\sum_{\sigma_0}\lambda_{\sigma_0}\langle\widetilde{x}(\sigma,s|x_0,\sigma_0)\rangle$ with the initial state probability $\lambda_{\sigma_0}$; from this quantity, by inverse Laplace transform, one infers the time evolution of the first moment on Fig. \ref{fig:x12}.


The stationary value of these first moments is retrieved by applying the final value theorem of the Laplace transform, i.e.,
\begin{equation}\label{eq:xst}
\begin{aligned}
\langle x \rangle_\sigma^{\rm st}&\equiv\lim_{t\rightarrow \infty} \langle x(\sigma,t|x_0,\sigma_0)\rangle=\lim_{s\rightarrow 0}s\, \langle\widetilde{x}(\sigma,s|x_0,\sigma_0)\rangle\\
&=\widetilde{\Psi}_\sigma(\nu)\langle x_S \rangle_\sigma^{\rm st} +\frac{c_\sigma}{2\langle \tau\rangle}\left[\langle \tau\rangle_\sigma-\widetilde{\Psi}_\sigma(\nu)\right]\\
&=\frac{\left[1-\widetilde{\psi}_+(\nu)\right]\left[1-\widetilde{\psi}_-(\nu)\right]}{2\nu\langle \tau\rangle}\left(c_{-\sigma}-c_\sigma\right)+\frac{c_\sigma\langle \tau\rangle_\sigma}{2\langle \tau\rangle},
\end{aligned}
\end{equation}
that yields the stationary average position
\begin{equation}\label{eq:xstat}
\langle x\rangle^{\rm st}=\frac{c_+\langle \tau\rangle_++c_-\langle \tau\rangle_-}{\langle \tau\rangle_++\langle \tau\rangle_-},
\end{equation}
which corresponds to the weighted average of the two centers $c_+$, and $c_-$ with respect to the corresponding average waiting time $\langle \tau\rangle_+$, and $\langle \tau\rangle_-$. 

\subsubsection{Second moment}

\begin{figure}
\centering
\includegraphics[width = 0.48\linewidth]{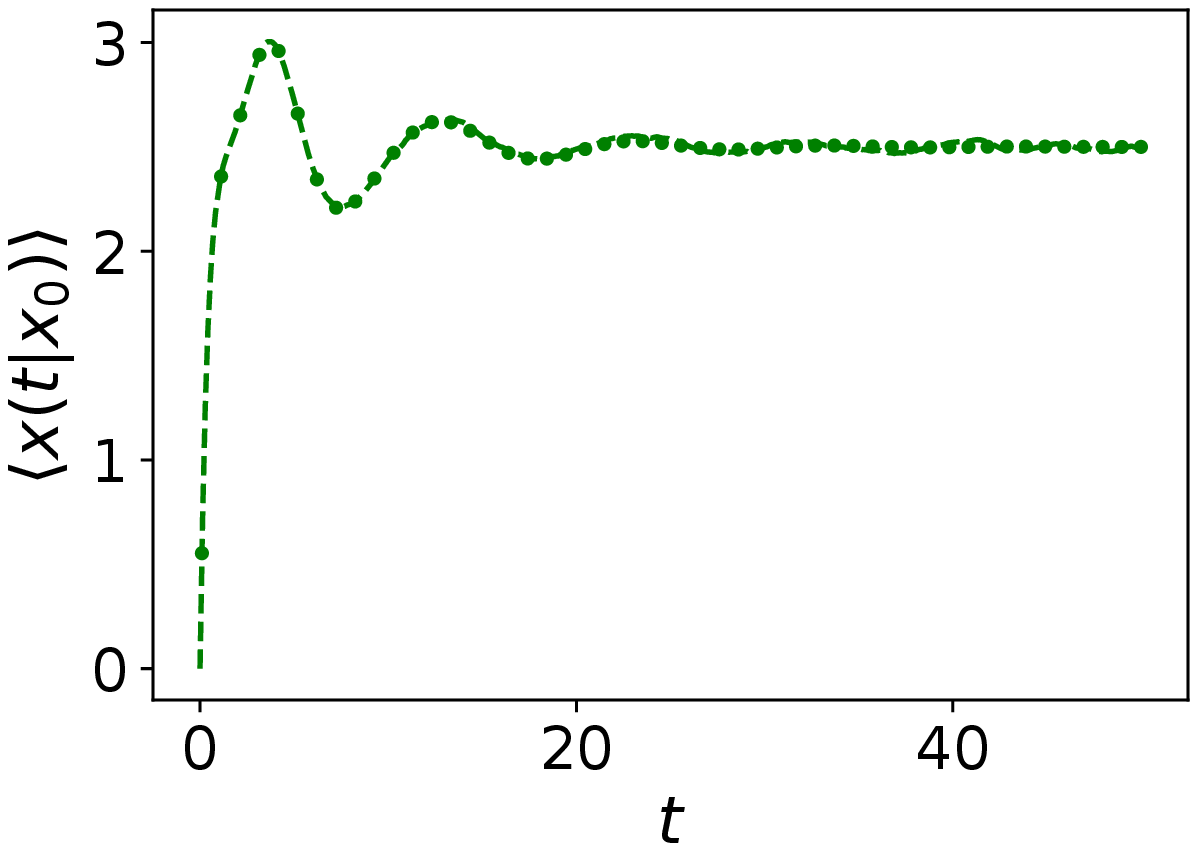}\quad \includegraphics[width = 0.48\linewidth]{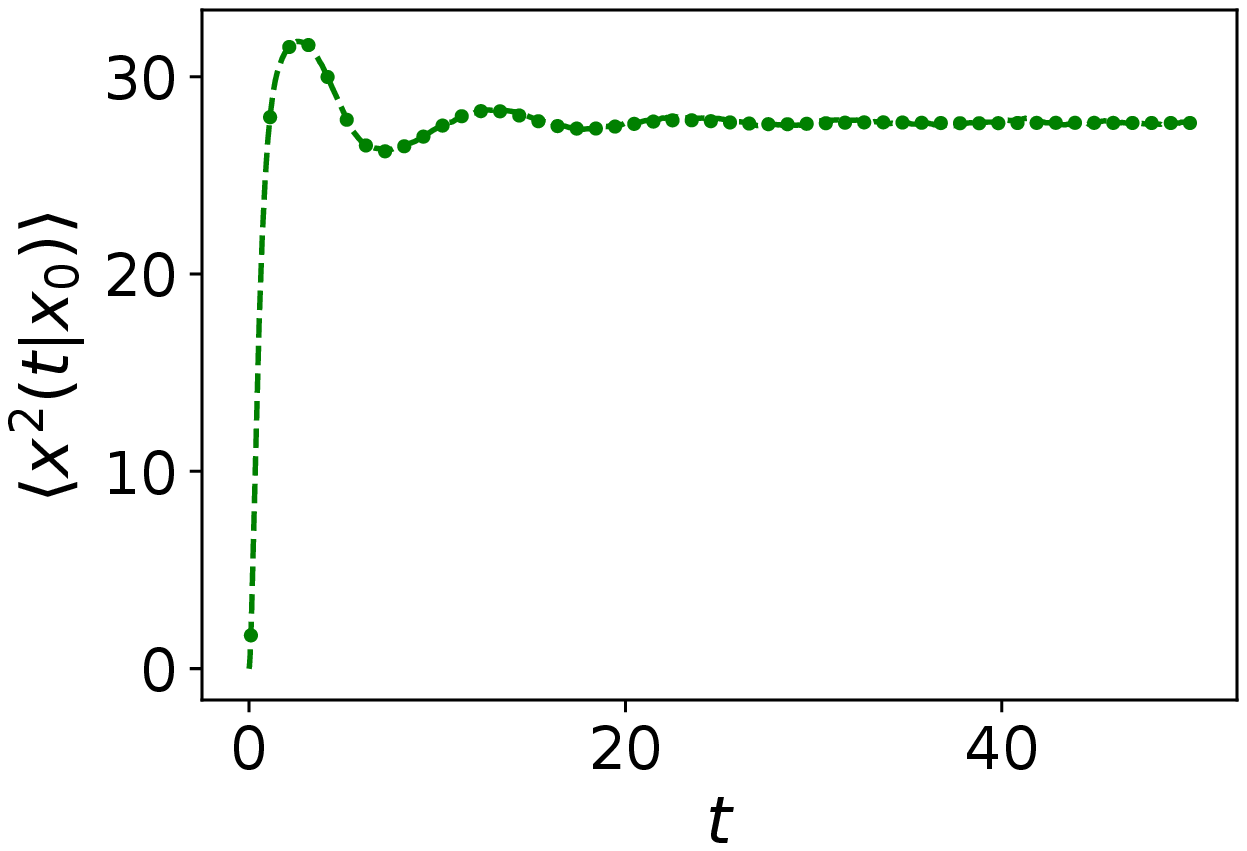}
\caption{Dependence of $\langle x(t|x_0)\rangle$ (left) and $\langle x^2(t|x_0)\rangle$ (right) on time $t$, for a particle starting at $x_0=0$ at $t=0$. In both pictures the dashed line corresponds to simulations ($N=10^5$ samples with $\Delta t=0.005$) and dots by inverse (numerical) Laplace transform of Eq. \eqref{eq:x2pp}. Due to the initial conditions, all curves start from $x_0=0$ at time $t=0$ but, after an oscillatory transient they reach their  stationary values in Eqs. \eqref{eq:xstat} and \eqref{eq:x2stat}. The parameters of the model are: $D=1$, $c_+=7.5$, $c_-=-2.5$, $\nu=2.5$, $k_+=10$, $\theta_+=0.5$, $k_-=5$, $\theta_-=1$, and the initial state probability $\lambda=0.5$.}\label{fig:x12}
\end{figure}

Analogously, we derive the second moments of the probability density $G(x,\sigma,t|x_0,\sigma_0)$ by multiplying Eq. \eqref{eq:GLT} by $x^2$ and integrating over $x$: 
\begin{figure}
\centering
\includegraphics[width = 0.48\linewidth]{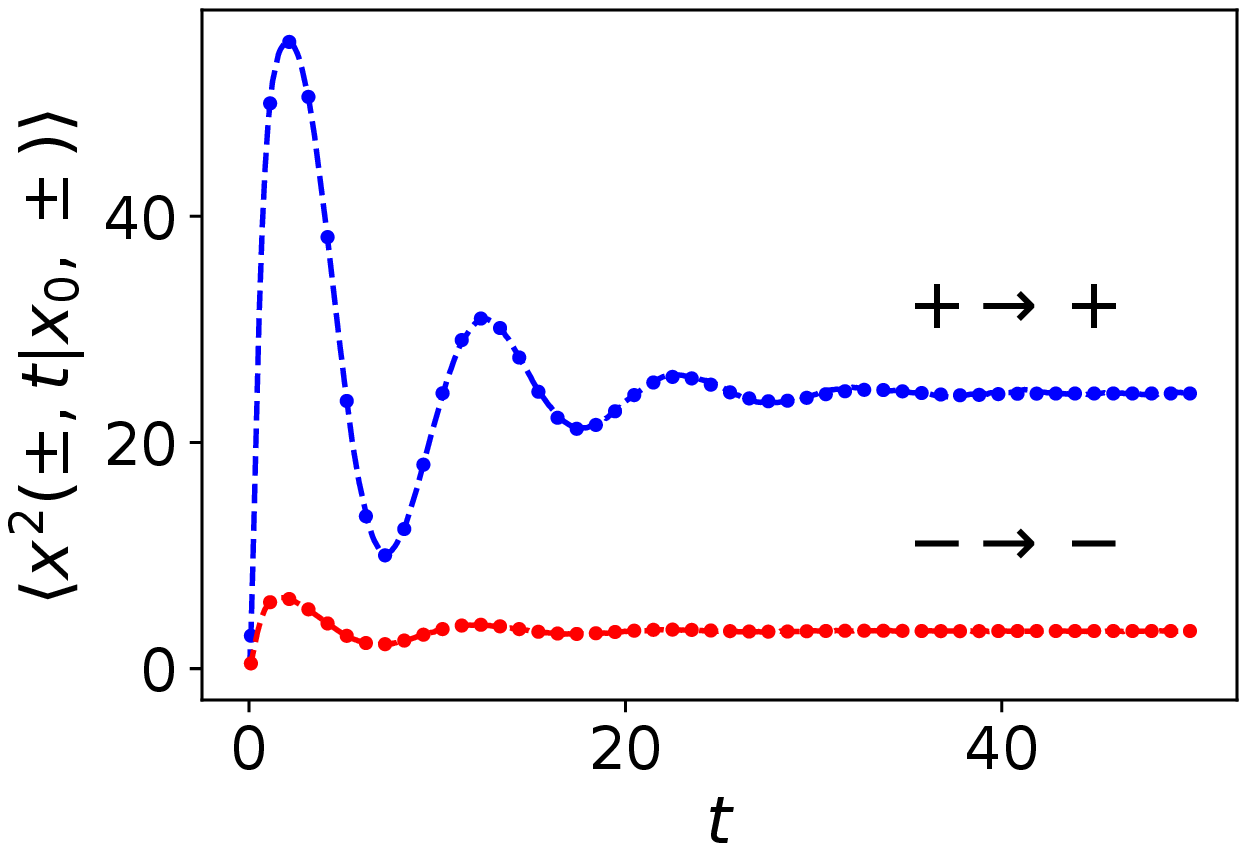}\quad \includegraphics[width = 0.48\linewidth]{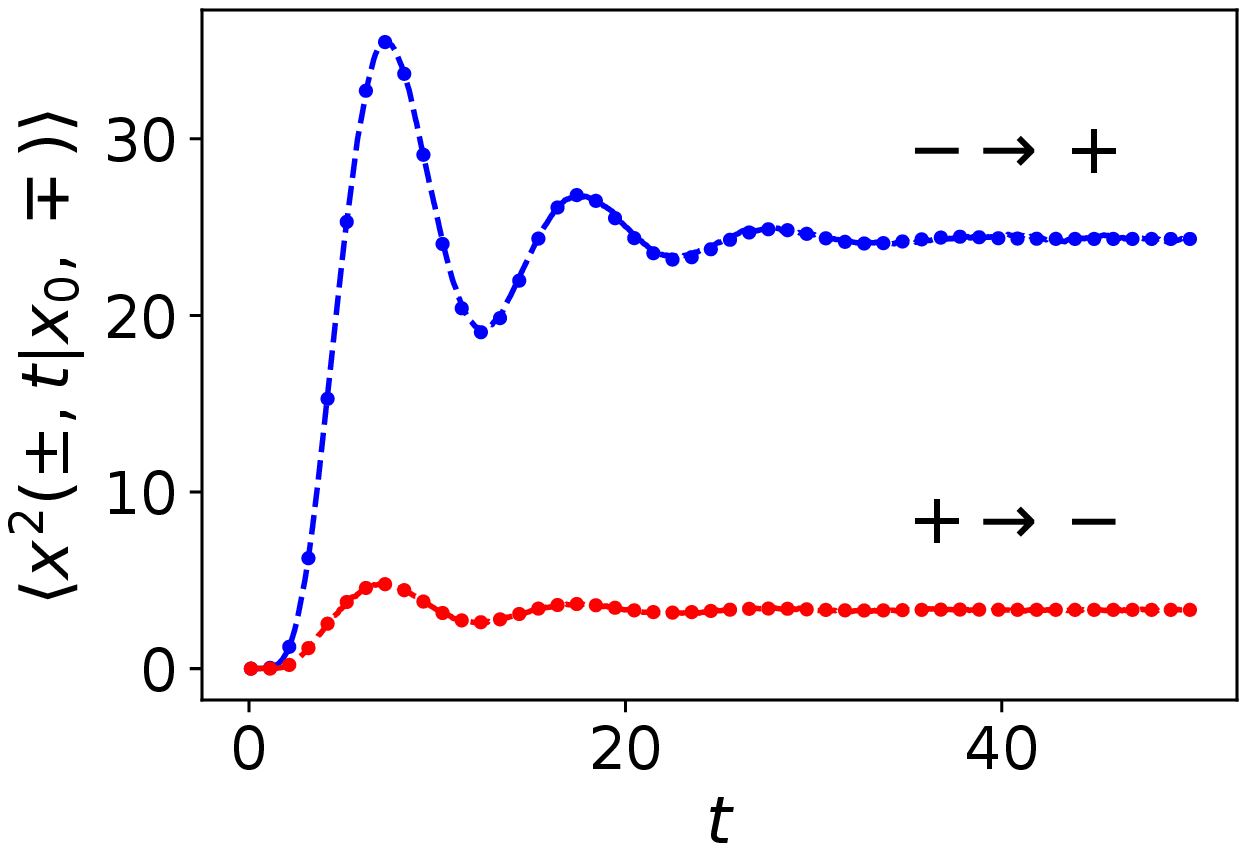}
\caption{Dependence of the conditional moments $\langle x^2(\sigma,t|x_0,\sigma_0)\rangle$ on time $t$ for the same (left) or different (right) initial and final potentials. In particular, in both panels the blue lines refer to a fixed potential centered in $c_+$, and the red ones to $c_-$, while  the dashed lines correspond to simulations ($N=10^5$ samples with $\Delta t=0.005$) and the dots to the  inverse Laplace transform of Eq. \eqref{eq:xpp}. Due to the initial conditions, all curves start from $x_0=0$ at time $t=0$ but, after an oscillatory transient they reach the stationary values given by Eqs. \eqref{eq:x2st}. The parameters of the model are: $D=1$, $c_+=7.5$, $c_-=-2.5$, $\nu=2.5$, $k_+=10$, $\theta_+=0.5$, $k_-=5$, $\theta_-=1$, $x_0=0$ and $\lambda=0.5$.}\label{fig:x2pp}
\end{figure}

\begin{equation}\label{eq:x2pp}
\begin{aligned}
\langle\widetilde{x}^2(\sigma,s|x_0,\sigma)\rangle=&\,\mathcal{L}\left\{\Psi_\sigma(t)\left[\Sigma^2(t)+\mu_\sigma^2(t|x_0)\right]\right\}(s)\\
&+\int_{-\infty}^{+\infty}\mathrm{d}y\,\widetilde{G}_S(y,\sigma,s|x_0,\sigma)\,\mathcal{L}\left\{\Psi_\sigma(t)\left[\Sigma^2(t)+\mu_\sigma^2(t|y)\right]\right\}(s)\\
=&\,x_0^2\,\widetilde{\Psi}_\sigma(s+2\nu)+2x_0\,c_\sigma\left[\widetilde{\Psi}_\sigma(s+\nu)-\widetilde{\Psi}_\sigma(s+2\nu)\right]\\
&+c_\sigma^2\left[\widetilde{\Psi}_\sigma(s)-2\widetilde{\Psi}_\sigma(s+\nu)+\widetilde{\Psi}_\sigma(s+2\nu)\right]+\frac{D}{\nu}\left[\widetilde{\Psi}_\sigma(s)-\widetilde{\Psi}_\sigma(s+2\nu)\right]\\
&+\langle\widetilde{x}_S^2(\sigma,s|x_0,\sigma)\rangle\,\widetilde{\Psi}_\sigma(s+2\nu)+2\langle\widetilde{x}_S(\sigma,s|x_0,\sigma)\rangle\,c_\sigma\left[\widetilde{\Psi}_\sigma(s+\nu)-\widetilde{\Psi}_\sigma(s+2\nu)\right]\\
&+\frac{\widetilde{\psi}_+(s)\widetilde{\psi}_-(s)}{1-\widetilde{\psi}_+(s)\widetilde{\psi}_-(s)}\left\{c_\sigma^2\left[\widetilde{\Psi}_\sigma(s)-2\widetilde{\Psi}_\sigma(s+\nu)+\widetilde{\Psi}_\sigma(s+2\nu)\right]+\frac{D}{\nu}\left[\widetilde{\Psi}_\sigma(s)-\widetilde{\Psi}_\sigma(s+2\nu)\right]\right\},\\
\langle\widetilde{x}^2(\sigma,s|x_0,-\sigma)\rangle=&\int_{-\infty}^{+\infty}\mathrm{d}y\,\widetilde{G}_S(y,\sigma,s|x_0,-\sigma)\,\mathcal{L}\left\{\Psi_\sigma(t)\left[\Sigma^2(t)+\mu_\sigma^2(t|y)\right]\right\}(s)\\
=&\,\widetilde{\Psi}_\sigma(s+\nu)\langle \widetilde{x}^2_S(\sigma,s|x_0,-\sigma)\rangle+ 2\,c_\sigma\,\langle \widetilde{x}_S(\sigma,s|x_0,-\sigma)\rangle\left[\widetilde{\Psi}_\sigma(s+\nu)-\widetilde{\Psi}_\sigma(s+2\nu)\right]\\
&+\frac{\widetilde{\psi}_{-\sigma}(s)}{1-\widetilde{\psi}_+(s)\widetilde{\psi}_-(s)}\left\{c_\sigma^2\left[\widetilde{\Psi}_\sigma(s)-2\widetilde{\Psi}_\sigma(s+\nu)+\widetilde{\Psi}_\sigma(s+2\nu)\right]+\frac{D}{\nu}\left[\widetilde{\Psi}_\sigma(s)-\widetilde{\Psi}_\sigma(s+2\nu)\right]\right\}.
\end{aligned}
\end{equation}
These quantities are plotted in Fig. \ref{fig:x2pp} for a representative choice of the various parameters.
Finally, the second moment of the position reads
$\langle\widetilde{x}^2(s|x_0)\rangle=\sum_\sigma \langle\widetilde{x}^2(\sigma,s|x_0)\rangle $, where $\langle \widetilde{x}^2(\sigma,s|x_0)\rangle=\sum_{\sigma_0}\lambda_{\sigma_0}\langle\widetilde{x}^2(\sigma,s|x_0,\sigma_0)\rangle$, from which, by Laplace transform, we infer the time evolution is displayed in the right panel of Fig. \ref{fig:x12}.

Always by direct application of the asymptotic theorem of the Laplace transform, we extract the stationary value of the second moment of the position conditioned on the final state $\sigma$, that is

\begin{equation}\label{eq:x2st}
\begin{aligned}
\langle x^2 \rangle_\sigma^{\rm st} =&\, \widetilde{\Psi}_\sigma(2\nu)\,\langle x^2_S\rangle_\sigma^{\rm st}+2c_\sigma\left[\widetilde{\Psi}_\sigma(\nu)-\widetilde{\Psi}_\sigma(2\nu)\right]\langle x_S\rangle_\sigma^{\rm st}\\ &+\frac{1}{2\langle \tau\rangle}\left\{c_\sigma^2 \left[\langle \tau\rangle_\sigma-\widetilde{\Psi}_\sigma(\nu)+\widetilde{\Psi}_\sigma(2\nu)\right]+\frac{D}{\nu}\left[\langle \tau\rangle_\sigma-\widetilde{\Psi}_\sigma(2\nu)\right]\right\},
\end{aligned}
\end{equation}
which immediately allows us to reconstruct the complete second moment as 

\begin{equation}\label{eq:x2stat}
  \langle x^2\rangle^{\rm st}=\langle x^2\rangle_+^{\rm st}+\langle x^2\rangle_-^{\rm st}.   
\end{equation}

\section{Statistics of $c(t)$}\label{App:B}

In this Section we determine again $P(\sigma,t)$ in Eq. \eqref{eq:Ps}, i.e., the probability that the process $c(t)$ takes the value $c_\sigma$ (with $\sigma\in\{\pm\}$) at time $t$, in a way that is more suitable for the calculation of the observables of interest. Moreover, this alternative analysis provides also an application of the with renewal approach. In order to access this probability distribution, it is useful to compute the conditional probability $P(\sigma_2,t_2|\sigma_1,t_1)$ of being in the state $\sigma_2$ at time $t_2$ given that the initial value $\sigma(t_1)=\sigma_1$ coincides with a switching event. It simply follows that $P(\sigma,t)=\sum_{\sigma_0}\lambda_{\sigma_0}P(\sigma,t|\sigma_0)$, where we omit the initial time $t_1$ whenever $t_1=0$.\\
The computation of $P(\sigma_2,t_2|\sigma_1,t_1)$ can be tackled by means of a renewal approach. Let us start from the evaluation of $P(+,t_2|+,t_1)$, which can be expressed as

\begin{equation}\label{eq:p++}
\begin{aligned}
P(+,t_2|+,t_1)&=\Psi_+(t_2-t_1)+\int_{t_1}^{t_2}\mathrm{d}\tau\,\psi_+(\tau-t_1)P(+,t|-,\tau)\\
&=\Psi_+(t_2-t_1)+\int_{0}^{t_2-t_1}\mathrm{d}\tau'\,\psi_+(\tau')P(+,t-\tau'|-)\\
&=P(+,t_2-t_1|+).
\end{aligned}
\end{equation}
The first term $\Psi_+(t_2-t_1)$ on the right hand side of Eq. \eqref{eq:p++} accounts for trajectories with no switches within the time interval $(t_1,t_2)$. The second contribution is computed conditioning on the first switch at time $\tau$, after which the process starts from the state $\sigma(\tau)=-$ to reach the final state $\sigma(t)=+.$
The second line in Eq. \eqref{eq:p++} simply follows from the change of variable $\tau'=t-\tau$, which makes apparent the time translation invariance of  $P(+,t_2|+,t_1)=P(+,t_2-t_1|+)$. This property holds for all the $P(\sigma_2,t_2|\sigma_1,t_1)$, which allows us to simply consider $t_1=0$ in all the calculations.
Similarly, the equations for the probabilities of all the other possible configurations of the final and initial state are given by

\begin{equation}\label{eq:peq}
\begin{aligned}
&P(\sigma,t|\sigma )=\Psi_\sigma(t)+\int_0^t\mathrm{d}\tau\,\psi_\sigma(\tau)P(\sigma,t-\tau|-\sigma),\\
&P(\sigma,t|-\sigma)=\int_0^t\mathrm{d}\tau\,\psi_\sigma(\tau)P(\sigma,t-\tau|\sigma ).
\end{aligned}
\end{equation}
It is natural to solve Eq. \eqref{eq:peq} in terms of Laplace transforms.
By applying the convolution theorem of the Laplace transform to Eqs. \eqref{eq:peq}, we find a closed expression for the Laplace transform $\widetilde{P}(\sigma,s|\pm \sigma)$:

\begin{equation}\label{eq:PsLT}
\begin{aligned}
&\widetilde{P}(\sigma  ,s|\sigma )=\frac{\widetilde{\Psi}_\sigma(s)}{1-\widetilde{\psi}_+(s)\widetilde{\psi}_-(s)},\\
&\widetilde{P}(\sigma  ,s|-\sigma )=\frac{\widetilde{\Psi}_\sigma(s)\widetilde{\psi}_{-\sigma}(s)}{1-\widetilde{\psi}_+(s)\widetilde{\psi}_-(s)},\\
\end{aligned}
\end{equation}
where we recall that the Laplace transform of the cumulative waiting-time distribution is  $\widetilde{\Psi}_\sigma(s)=[1-\widetilde{\psi}_\sigma(s)]/s$.
Note that Eqs.~\eqref{eq:PsLT} coincide exactly with Eqs. \eqref{eq:Psc}, and they satisfy the normalisation condition expressed for the Laplace transforms, e.g.,  $\widetilde{P}(\sigma,s|\sigma_0)+\widetilde{P}(-\sigma,s|\sigma_0)=1/s$ for any $\sigma_0$.

\subsection{General Probability Distribution with asymmetric waiting times}
In this Section we generalise the probability $P(\sigma,t|\sigma_0)$ discussed above by computing the probability to reach a state $\sigma_2$ at $t_2$ starting from the state $\sigma_1$ at a time $t_1$, not necessarily corresponding to an initial switching event. We denote this probability by $P_{\sigma_0}(\sigma_2,t_2|\sigma_1,t_1)$, with $\sigma_0,\,\sigma_1,\,\sigma_2\in\{-,+\}$, being the probability that the process reaches $\sigma(t_2)=\sigma_2$ from $\sigma(t_1)=\sigma_1$ given the initial condition $\sigma(0)=\sigma_0$. 
\paragraph{Conditional probability $P_+(+,t_2|+,t_1)$:}
We start by deriving an integral equation for $P_+(+,t_2|+,t_1)$, that we can express, via the renewal approach, as:
\begin{equation}\label{eq:p+++}
\begin{aligned}
P_+(+,t_2|+,t_1)=&\Psi_+(t_2)+\int_{t_1}^{t_2}\mathrm{d}\tau\,\psi_+(\tau)P(+,t_2-\tau|-)\\
&+\int_0^{t_1}\mathrm{d}\tau_1\int_{t_1}^{t_2}\mathrm{d}\tau_2\,\psi_+(\tau_2-\tau_1)P(+,t_2-\tau_2|-)P_S(+,\tau_1|+)\\
&+\int_0^{t_1}\mathrm{d}\tau_1\int_{t_2}^{\infty}\mathrm{d}\tau_2\,\psi_+(\tau_2-\tau_1)P_S(+,\tau_1|+)\\
=&\Psi_+(t_1+t)+\int_{0}^{t}\mathrm{d}\tau'\psi_+(t+t_1-\tau')P(+,\tau'|-)\\
&+\int_0^{t_1}\mathrm{d}\tau_1'\int_{0}^{t}\mathrm{d}\tau_2'\,\psi_+(t-\tau_2'+\tau_1')P(+,\tau_2'|-)P_S(+,t-\tau_1'|+)\\
&+\int_0^{t_1}\mathrm{d}\tau_1\int_{0}^{\infty}\mathrm{d}\tau_2'\,\psi_+(t+t_1+\tau_2'-\tau_1)P_S(+,t-\tau_1'|+).\\
\end{aligned}
\end{equation}
The first contribution $\Psi_+(t_2)$ on the right hand side of Eq.~\eqref{eq:p+++} describes the case where $\sigma(t)=+$ for the whole time interval $(0,t_2)$, that is, no switching event occurs up to time $t_2$. The second contribution considers the case in which the first switching event happens at a time $\tau\in(t_1,t_2)$, with probability $\psi_+(\tau)$; then the process reaches the state $+$ at time $t_2$ from the state $-$ at the switching time $\tau$, which comes with a probability weight $P(+,t_2-\tau|-)$. The second line corresponds to the case in which a switch has occurred before $t_1$ while, in the interval $(\tau_1,\tau_2)$ to which $t_1$ belongs, no switch occurs, so that the state is fixed to $+$, contributing with probability $\psi_+(\tau_2-\tau_1)$. In order to have a $+$ state in the interval $(\tau_1,\tau_2)$, being $\sigma(0)=+$, an even number of switches must take place in the interval $(0,\tau_1]$, whose probability is given by $P_S(+,\tau_1|+)$ in Eq. \eqref{eq:PSt}; finally, the process attains the state $\sigma(t_2)=+$ from the state $\sigma(\tau_2)=-$ with probability $P(+,t_2-\tau_2|-)$. The third line represents the case in which the last switch before $t_1$ happens at time $\tau_1$, and than no switching occurs in the interval $(\tau_1,t_2)$. The second inequality follows from the shift of the integration variables, such that the dependence on the time $t_1$ and $t\equiv t_2-t_1$ is apparent. Indeed, our ultimate goal is to compute the stationary limit of Eq.~\eqref{eq:p+++}, corresponding to $t_1\rightarrow\infty$ and keeping $t$ fixed.
In order to simplify the calculations, we compute the Laplace transform of Eq. \eqref{eq:p+++} with respect to $t_1$, which reads

\begin{equation}\label{eq:p+++t1}
\begin{aligned}
\widetilde{P}_+(+,t|+,\eta)\equiv\mathcal{L}_{t_1}\left\{P_+(+,t+t_1|+,t_1)\right\}(\eta)=&\int_0^\infty\mathrm{d}t_1\,e^{-\eta t_1}\left[\int_{t_1+t}^{\infty}\mathrm{d}\tau\,\psi_+(\tau)+\int_{0}^{t}\mathrm{d}\tau\,\psi_+(t+t_1-\tau)P(+,\tau|-)\right]\\
&+\widetilde{P}_S(+,\eta|+)\int_{0}^{t}\mathrm{d}\tau\,P(+,\tau|-)\int_0^\infty\mathrm{d}y\,e^{-y\eta}\psi_+(t-\tau+y)\\
&+\widetilde{P}_S(+,\eta|+)\int_{0}^{\infty}\mathrm{d}\tau\,\int_0^\infty\mathrm{d}y\,e^{-y\eta}\psi_+(t+\tau+y),
\end{aligned}
\end{equation}
where the convolution theorem of the Laplace transform allows us to factorize the contribution of $\widetilde{P}_S(+,\eta|+)$ in the second and third line, whose expression is reported in Eq. \eqref{eq:norms}.
The convergence of $\widetilde{P}_S(+,\eta|+)$ follows from the condition $|\psi(\eta)|<1$ for $\operatorname{Re}(\eta)>0$, while, as we will see, the final expression can be analytically extended to $\eta=0$, consistently with the final value theorem of the Laplace transform.

We can further simplify Eq. \eqref{eq:p+++t1} by taking the Laplace transform with respect to $t$:

\begin{equation}\label{eq:p+++t}
\begin{aligned}
\widetilde{P}_+(+,s|+,\eta)\equiv\mathcal{L}_{t}\left\{\widetilde{P}_+(+,t|+,\eta)\right\}(s)=&[1+\widetilde{P}_S(+,\eta|+)]\left[\widetilde{P}(+,s|-)\psi_d^+(\eta,s)+\Psi_d^+(\eta,s)\right].
\end{aligned}
\end{equation}
In the equation above, we have introduced the incremental ratios:

\begin{equation}
\psi_d^\sigma(\eta,s)\equiv\frac{\widetilde{\psi}_\sigma(\eta)-\widetilde{\psi}_\sigma(s)}{s-\eta},\,\,\,\,\,\,\Psi_d^\sigma(\eta,s)\equiv\frac{\widetilde{\Psi}_\sigma(\eta)-\widetilde{\Psi}_\sigma(s)}{s-\eta},
\end{equation}
corresponding to the joint Laplace transform of integrals of the type
\begin{equation}
f_d(\eta,s)\equiv\int_0^\infty \mathrm{d}\tau_1\,e^{-\eta \tau_1}\int_0^\infty \mathrm{d}\tau_2\,e^{-s \tau_2}\,f(\tau_1+\tau_2)=\frac{\widetilde{f}(\eta)-\widetilde{f}(s)}{s-\eta},
\end{equation}
where $f$ is any function for which its Laplace transform $\widetilde{f}$ is well defined.

In order to compute the stationary limit of Eq.~\eqref{eq:p+++t}, we assume that $\psi(t)$ displays finite first moment ${\langle\tau\rangle\equiv\int_0^\infty\mathrm{d}\tau\,\tau\psi(\tau)}$. Hence, we can expand its Laplace transform $\widetilde{\psi}_\sigma(\eta)$ around $\eta=0$ as $\widetilde{\psi}_\sigma(\eta)=1-\eta\langle \tau\rangle_\sigma+o(\eta)$.  
Accordingly, Eq. \eqref{eq:p+++t} allows us to compute the stationary limit $\widetilde{P}_{\rm st}(+,s|+)$ of the distribution by means of the final value theorem of the Laplace transform, that is

\begin{equation}\label{eq:P++Laplace transform}
\widetilde{P}_{\rm st}(+,s|+)=\lim_{\eta\rightarrow 0} \eta \widetilde{P}_+(+,s|+,\eta)=\frac{1}{2\langle \tau\rangle}\left[\frac{\langle \tau\rangle_+}{s}-\frac{\widetilde{\Psi}_+(s)\widetilde{\Psi}_-(s)}{1-\widetilde{\psi}_+(s)\widetilde{\psi}_-(s)}\right],
\end{equation}
that follows from the expressions

\begin{equation}
\lim_{\eta\rightarrow 0}\eta \left[1+\widetilde{P}_S(+,s|+)\right]=\frac{1}{2\langle \tau\rangle}\,\,\,\,\,\text{and}\,\,\,\,\,\lim_{\eta\rightarrow 0}\widetilde{\Psi}_\sigma(\eta)=\langle \tau\rangle_\sigma,
\end{equation}
with $\langle \tau\rangle$ given by Eq.~\eqref{eq:avtau}.

It can be easily checked that $P_-(-,t_2|-,t_1)$ satisfies the same equation \eqref{eq:p+++} that  $P_+(+,t_2|+,t_1)$ satisfies where all the $+$ states are replaced by $-$ ones and vice versa. This reasoning generalizes to all the configurations.

\paragraph{Conditional probability $P_+(+,t_2|-,t_1)$:}
Following the same steps as above, we can express the conditional probability $P_+(+,t_2|-,t_1)$ as
\begin{equation}\label{eq:p+-+}
\begin{aligned}
P_+(+,t_2|-,t_1)&=
\int_0^{t_1}\mathrm{d}\tau_1\int_{t_1}^{t_2}\mathrm{d}\tau_2\,\psi_-(\tau_2-\tau_1)P(+,t_2-\tau_2|+)P_S(-,\tau_1|+)\\
&=\int_0^{t_1}\mathrm{d}\tau_1'\int_{0}^{t}\mathrm{d}\tau_2'\,\psi_-(t-\tau_2'+\tau_1')P(+,\tau_2'|+)P_S(-,t_1-\tau_1'|+),
\end{aligned}
\end{equation}
where the right-hand side of the equation accounts for trajectories that (i) have an odd number of switches in the interval $(0,\tau_1]$ before $t_1$, weighted by $P_S(-,\tau_1|+)$, (ii) have no switching in the interval $(\tau_1,\tau_2)$, with $\tau_2\in(t_1,t_2)$, with probability $\psi_-(\tau_2-\tau_1)$, (iii) reach the state $\sigma(t_2)=+$ from $\sigma(\tau_2)=+$ according to $P(+,t_2-\tau_2|+).$ The second line of the equation follows from the change of variables $\tau_2'=t_2-\tau_2$, $\tau_1'=t_1-\tau_1$ and $t=t_2-t_1$, and allows us to apply  the convolution theorem of the Laplace transform.
The Laplace transform of Eq. \eqref{eq:p+-+} with respect to $t_1$, with conjugate variable $\eta$, is given by

\begin{equation}\label{eq:p+-+t1}
\widetilde{P}_+(+,t_2|-,\eta)\equiv\mathcal{L}_{t_1}\left\{P_+(+,t_2|-,t_1)\right\}(\eta)=\widetilde{P}_S(-,\eta|+)\int_0^t\mathrm{d}\tau\,P(+,\tau|+)\int_0^\infty\mathrm{d}y\,e^{-\eta y}\psi_-(t-\tau+y),
\end{equation}
where the Laplace of $P_S(-,t|+)$ is given by Eq. \eqref{eq:norms}.
Similarly to what was done previously, we compute the Laplace transform $\widetilde{P}_+(+,s|-,\eta)$ also with respect to $t$, i.e.,

\begin{equation}\label{eq:p+-+t}
\widetilde{P}_+(+,s|-,\eta)\equiv\mathcal{L}_{t}\left\{\widetilde{P}_+(+,t|-,\eta)\right\}(s)=\widetilde{P}_S(-,\eta|+)\widetilde{P}(+,s|+)\psi_d^-(\eta,s).
\end{equation}
The Laplace transform of the stationary state probability $\widetilde{P}(+,s|-)$ can be immediately computed from Eq. \eqref{eq:p+-+t} by applying the final value theorem of the Laplace transform, i.e.,

\begin{equation}\label{eq:p+-}
\widetilde{P}_{\rm st}(+,s|-)=\lim_{\eta\rightarrow 0}\eta \widetilde{P}_+(+,s|-,\eta)=\frac{\widetilde{P}(+,s|+)}{2\langle \tau\rangle}\widetilde{\Psi}_-(s)=\frac{1}{2\langle \tau\rangle }\frac{\widetilde{\Psi}_-(s)\widetilde{\Psi}_+(s)}{1-\widetilde{\psi}_-(s)\widetilde{\psi}_+(s)}.
\end{equation}

\paragraph{Conditional probability $P_+(-,t_2|+,t_1)$:}
The conditional probability $P_+(-,t_2|+,t_1)$ can be written as the sum of two contributions:
\begin{equation}\label{eq:p++-}
\begin{aligned}
P_+(-,t_2|+,t_1)=&\int_{t_1}^{t_2}\mathrm{d}\tau\,\psi_+(\tau)P(-,t_2-\tau|-)\\
&+\int_0^{t_1}\mathrm{d}\tau_1\int_{t_1}^{t_2}\mathrm{d}\tau_2\,\psi_+(\tau_2-\tau_1)P(-,t_2-\tau_2|-)P_S(+,\tau_1|+)\\
=&\int_{0}^{t}\mathrm{d}\tau'\,\psi_+(t+t_1-\tau')P(-,\tau'|-)\\
&+\int_0^{t_1}\mathrm{d}\tau_1'\int_{0}^{t}\mathrm{d}\tau_2'\,\psi_+(t-\tau_2'+\tau_1')P(-,\tau_2'|-)P_S(+,t_1-\tau_1'|+).
\end{aligned}
\end{equation}
The first contribution comes from trajectories associated with a first switching at $\tau\in(t_1,t_2)$; the second, to trajectories characterized by the occurrence of any even number of switches before $t_1$. The second equality comes from a simple change of variables.
Following the same steps as in the  previous Sections, we can express the Laplace transform of $P_+(-,t_2|+,t_1)$ with respect to $t_1$ and $t$, denoted by $\widetilde{P}_+(-,s|+,\eta)$, as

\begin{equation}
\widetilde{P}_+(-,s|+,\eta)=\left[1+\widetilde{P}_S(+,\eta|+)\right]\widetilde{P}(-,s|-)\psi_d^+(\eta,s).
\end{equation}



The stationary limit of Eq. \eqref{eq:p++-} gives the stationary probability $P_{\rm st}(-,t|+)$, whose Laplace transform is computed to be 


\begin{equation}\label{eq:P-+Laplace transform}
\widetilde{P}_{\rm st}(-,s|+)=\lim_{\eta\rightarrow 0}\eta \widetilde{P}_+(-,s|+,\eta)=\frac{\widetilde{P}(-,s|-)}{2\langle \tau\rangle}\widetilde{\Psi}_+(s)=\frac{1}{2\langle \tau\rangle }\frac{\widetilde{\Psi}_-(s)\widetilde{\Psi}_+(s)}{1-\widetilde{\psi}_-(s)\widetilde{\psi}_+(s)}.
\end{equation}

\paragraph{Conditional probability $P_+(-,t_2|-,t_1)$:}
The conditional probability $P_+(-,t_2|-,t_1)$ also comes with two contributions:
\begin{equation}
\begin{aligned}
P_+(-,t_2|-,t_1)=&\int_0^{t_1}\mathrm{d}\tau_1\int_{t_1}^{t_2}\mathrm{d}\tau_2\,\psi_-(\tau_2-\tau_1)P(-,t_2-\tau_2|+)P_S(-,\tau_1|+)\\
&+\int_0^{t_1}\mathrm{d}\tau_1\int_{t_2}^{\infty}\mathrm{d}\tau_2\,\psi_-(\tau_2-\tau_1)P_S(-,\tau_1|+).
\end{aligned}
\end{equation}
The integrals account for trajectories with an odd number of switches before $t_1$ and at least one in $(t_1,t_2)$, which determine the first line, or none, which give the second line.
In this case we have


\begin{equation}
\widetilde{P}_+(-,s|-,\eta)=\widetilde{P}_S(-,\eta|+)\left[\widetilde{P}(-,s|+)\psi_d(\eta,s)+\Psi_d^-(\eta,s)\right],
\end{equation}
whose stationary limit is given by
\begin{equation}
\widetilde{P}_{\rm st}(-,s|-)=\lim_{\eta\rightarrow 0}\eta \widetilde{P}_+(-,s|-,\eta)=\frac{1}{2\langle \tau\rangle}\left[\frac{\langle \tau\rangle_-}{s}-\frac{\widetilde{\Psi}_+(s)\widetilde{\Psi}_-(s)}{1-\widetilde{\psi}_+(s)\widetilde{\psi}_-(s)}\right].
\end{equation}
Finally, we check the Laplace transform normalization condition of $\widetilde{P}_{\sigma_0}(\sigma_2,s|\sigma_1,\eta)$, which is equal 
\begin{equation}
\sum_{\sigma_2,\sigma_1\in\{\pm\}}\widetilde{P}_{\sigma_0}(\sigma_2,s|\sigma_1,\eta)=\frac{1}{s\eta}.
\end{equation}

\subsection{Correlator and power spectrum}

In this Section, we determine the two-point stationary correlator $C(t)$ of the process $c(t)$: 

\begin{equation}\label{eq:Ct}
\begin{aligned}
C(t)&\equiv\lim_{\tau\rightarrow\infty}\langle c(t+\tau)c(\tau)\rangle\\
&=\lim_{\tau\rightarrow\infty}\langle \left[c_s(t+\tau)+c_m\right]\left[c_s(\tau)+c_m\right]\rangle\\
&=\lim_{\tau\rightarrow\infty}\left\{\langle c_s(t+\tau)c_s(\tau)\rangle+c_m\left[\langle c_s(t+\tau)\rangle+\langle c_s(\tau)\rangle\right]+c_m^2\right\},
\end{aligned}
\end{equation}
where we denote by $\langle\cdots\rangle$ the expectation value over configurations of the process, and we have decomposed the process $c(t)$ in its symmetrized part $c_s(t)\in\{\pm c_0\}$, with 
\begin{equation}\label{eq:c0}
c_0\equiv (c_+-c_-)/2, 
\end{equation}
and its asymmetric contribution deriving from the average point
\begin{equation}\label{eq:cm}
c_m\equiv (c_++c_-)/2.
\end{equation}
We can compute $C(t)$ in Eq.~\eqref{eq:Ct} by casting an explicit expression of the expectation value $\langle\cdots\rangle$ with respect to the probability $P_{\sigma_0}(\sigma_2,t+\tau|\sigma_1, \tau)$ of the trajectories computed in the previous Section, and then by taking the stationary limit. Let us start by calculating the symmetric contribution to $C(t)$:
\begin{equation}
\begin{aligned}
C_s(t)=&c_0^2\lim_{\tau\rightarrow\infty}\sum_{\sigma_2,\sigma_1, \sigma_0}\lambda_{\sigma_0} \sigma_1 \sigma_2    \,P_\chi(\sigma_2,t+\tau|\sigma_1,\tau)
\\
=&c_0^2\left[P_{\rm st}(+,t|+)+P_{\rm st}(-,t|-)-2P_{\rm st}(-,t|+)\right],
\end{aligned}
\end{equation}
where the specific value of $\lambda_{\sigma_0}$ does not affect the stationary probability, and where the second equality results from the fact that in the stationary limit $P_{\rm st}(+,t|-)=P_{\rm st}(-\sigma,t|+\sigma)$. We can use results in Eqs. \eqref{eq:P++Laplace transform} and \eqref{eq:P-+Laplace transform} from the previous Section in order  to compute the Laplace transform $\widetilde{C}(s)$, i.e.,

\begin{equation}\label{eq:Cst}
\begin{aligned}
\widetilde{C}_s(s)=&c_0^2\left[\widetilde{P}_{\rm st}(+,s|+)+\widetilde{P}_{\rm st}(-,s|-)-2\widetilde{P}_{\rm st}(-,s|+)\right]\\
=&c_0^2\left[\frac{1}{s}-\frac{2}{\langle \tau\rangle}\frac{\widetilde{\Psi}_-(s)\widetilde{\Psi}_+(s)}{1-\widetilde{\psi}_-(s)\widetilde{\psi}_+(s)}\right].\\
\end{aligned}
\end{equation}

\begin{figure}
\centering
\includegraphics[width = 0.6\linewidth]{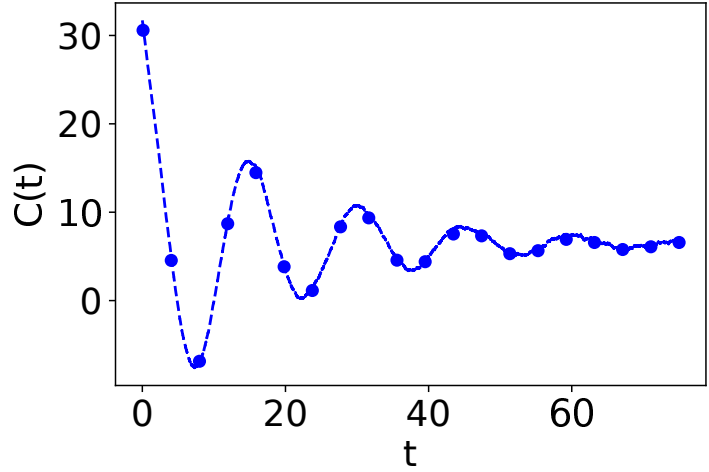}
\caption{Time evolution of the stationary autocorrelator $C(t)$ of the process $c(t)$. The dashed line corresponds to simulations ($N=10^5$ samples with $\Delta t=0.001$) while the dots to inverse Laplace transform of Eq. \eqref{eq:CLT}. The parameters of the model are: $D=1$, $c_+=7.5$, $c_-=-2.5$, $\nu=2.5$, $k_+=15$, $\theta_+=0.5$, $k_-=10$, $\theta_-=0.75$, $x_0=0$ and $\lambda=0.5$.}\label{fig:C}
\end{figure}

The asymmetric contribution to $C(t)$ is given by

\begin{equation}\label{eq:Ca}
\begin{aligned}
C_a(t)&=c_m^2+2c_m\lim_{\tau\rightarrow\infty}\langle c_s(\tau)\rangle\\
&=c_m^2+2c_m c_0\lim_{s\rightarrow 0} s \left[\widetilde{P}(+,s)-\widetilde{P}(-,s)\right]\\
&=c_m^2+2c_m c_0\frac{\langle \tau\rangle_+-\langle \tau\rangle_-}{\langle \tau\rangle_++\langle \tau\rangle_-},
\end{aligned}
\end{equation}
where we have used the exact expression of $\widetilde{P}(\sigma,s)$ in Eq.~\eqref{eq:Ps}, whose Laplace transform is $\widetilde{C}_a(s)=C_a/s.$
Finally, combining Eqs. \eqref{eq:Cst} and \eqref{eq:Ca}, we get the Laplace transform of $C(t)$, i.e.,
\begin{equation}\label{eq:CLT}
\widetilde{C}(s)=\widetilde{C}_s(s)+\widetilde{C}_a(s)=c_0^2\left[\frac{1}{s}-\frac{2}{\langle \tau\rangle}\frac{\widetilde{\Psi}_-(s)\widetilde{\Psi}_+(s)}{1-\widetilde{\psi}_-(s)\widetilde{\psi}_+(s)}\right]+\frac{1}{s}\left[c_m^2+2c_m c_0\frac{\langle \tau\rangle_+-\langle \tau\rangle_-}{\langle \tau\rangle_++\langle \tau\rangle_-}\right];
\end{equation}
from this equation, by inverse Laplace transform, one infers the time evolution of $C(t)$ on Fig. \ref{fig:C}.

Despite the appearence of a prefactor $1/s$ in front of Eq. \eqref{eq:CLT}, $\widetilde{C}(s)$ can be analytically continued to $s=0$ with $\widetilde{C}(0)=0$, and hence on the whole imaginary axis. Indeed, this is consistent with fact that at large times the correlator decays to zero, i.e.,  $\lim_{t\rightarrow\infty}C(t)=\lim_{s\rightarrow 0}s\widetilde{C}(s)=0.$ 
The fact that $\widetilde{C}(s)$ is analytic on the imaginary axis and the symmetry property $C(t)=C(-t)$ allows us to extract the Fourier transform $\mathcal{F}\left\{C(t)\right\}(\omega)$ of $C(t)$ by the simple relation

\begin{equation}
\mathcal{F}\left\{C\right\}(\omega)=\widetilde{C}(i\omega)+\widetilde{C}(-i\omega).
\end{equation}
We also have, by Wiener-Khinchin theorem \cite{van1992stochastic}, that the power spectrum $S_c(\omega)$ for $c(t)$, equates the Fourier Transform of $C(t)$,  that is

\begin{equation}\label{eq:pws}
S_c(\omega)=\widetilde{C}(i\omega)+\widetilde{C}(-i\omega).
\end{equation}
Equation \eqref{eq:pws} is very general and requires only the non-divergence of the moments of the waiting-time distribution $\psi(t)$. In particular, for the Gamma distribution we find Eq.~\eqref{eq:psd}.
%
Equation \eqref{eq:psd} are independent of $c_m$ because the power dissipated by the system depends only on the relative excursion of the $x(t)$ whenever a switch occurs.

\section{Stationary power}\label{App:C}

In this Section we show how to relate the expressions 
of the switching probability density $G_S(x,\sigma,t|x_0,\sigma_0)$ in Eqs. \eqref{eq:xspp}, \eqref{eq:xspm},  
to the first moment of the work $W(t).$ We start by introducing the work $\mathrm{d}W(t)$ done in the infinitesimal time interval $(t,t+\mathrm{d}t)$, which can be expressed via the change of the potential energy $V(t)=\kappa [x(t)-c(t)]^2/2$ due to the variation of the external stochastic parameter $c(t)$:

\begin{equation}\label{eq:work}
\begin{aligned}
\mathrm{d}W(t)&=\frac{\partial V}{\partial c}\circ \mathrm{d}c(t)=-\kappa\left[x(t)-c(t)\right]\circ \mathrm{d}c(t)\\
&=-\frac{\kappa}{2}\left[x(t+\mathrm{d}t)+x(t)-2c_m\right]\left[c_s(t+\mathrm{d}t)-c_s(t)\right],
\end{aligned}
\end{equation}
where in the last line we make explicit the Stratonovich product $\circ$ by the introduction of $c_s(t)$, the symmetrized version of the process $c(t)$ that takes value $\pm c_0$ with $c_0$ in Eq. \eqref{eq:c0}, and $c_m$ in Eq. \eqref{eq:cm}.

We start by computing the probability density $P_{\mathrm{d}W}(w|t)$ of $\mathrm{d}W(t)$:

\begin{equation}\label{eq:Pdw}
\begin{aligned}
P_{\mathrm{d}W}(w|t)=&\delta(w)\left\{ 1-\mathrm{d}t\left[\lambda \left(P_S(-,t|+)+P_S(+,t|+)\right)+(1-\lambda)\left(P_S(-,t|-)+P_S(+,t|-)\right)\right]\right\}\\
&+\mathrm{d}t\int_{-\infty}^{+\infty}\mathrm{d}x\left\{  \delta\left(w-2\kappa c_0 (x-c_m)\right)\left[\lambda\, G_S(x,-,t|x_0,+)+(1-\lambda)G_S(x,-,t|x_0,-)\right]\right.\\
&\;\;\;\;\;\;\;\;\;\;\;\;\;\;\;\;\;\;\;+\left. \delta\left(w+2\kappa c_0 (x-c_m)\right)\left[\lambda\, G_S(x,+,t|x_0,+)+(1-\lambda)G_S(x,+,t|x_0,-)\right] \right\}.
\end{aligned}
\end{equation}
The first line in Eq. \eqref{eq:Pdw} accounts for trajectories that do not display a switch at time $t$, yielding $\mathrm{d}W(t)=0$. The second and the third lines, instead, display respectively trajectories that switch from $c_\pm$ to $c_\mp$ at time $t$, such that the work done in $(t,t+\mathrm{d}t)$ is given by $\mathrm{d}W(t)=\pm2\kappa c_0 (x-c_m)$. 
Thanks to the expression of  $P_{\mathrm{d}W}(w|t)$ in Eq. \eqref{eq:Pdw}, we find its moments

\begin{equation}\label{eq:dwn}
\begin{aligned}
\langle\left(\mathrm{d}W(t)\right)^n\rangle=(2\kappa c_0)^n\mathrm{d}t\int_{-\infty}^{+\infty}\mathrm{d}&x \,(x-c_m)^n \left[\rho_-^S(x,t|x_0)+(-1)^n\,\rho_+^S(x,t|x_0)\right],
\end{aligned}
\end{equation}
where we consider $n\ge 1$, since for $n=0$ we get the normalization condition, and we have introduced the probability densities $\rho_\sigma^S(x,t|x_0)\equiv\sum_{\sigma_0} \lambda_{\sigma_0} G_S(x,\sigma,t|x_0,\sigma_0)$, and $\rho^S(x,t|x_0)\equiv\rho_+^S(x,t|x_0)+\rho_-^S(x,t|x_0)$. Note that, independently of the degree $n$ of the moment, $\langle\left(\mathrm{d}W(t)\right)^n\rangle$ is always proportional to $\mathrm{d}t$, such that integrals its time integrals are not infinitesimal, as one would have for the Wiener process \cite{gardiner}.

\begin{figure}
\centering
\includegraphics[width = 0.6\linewidth]{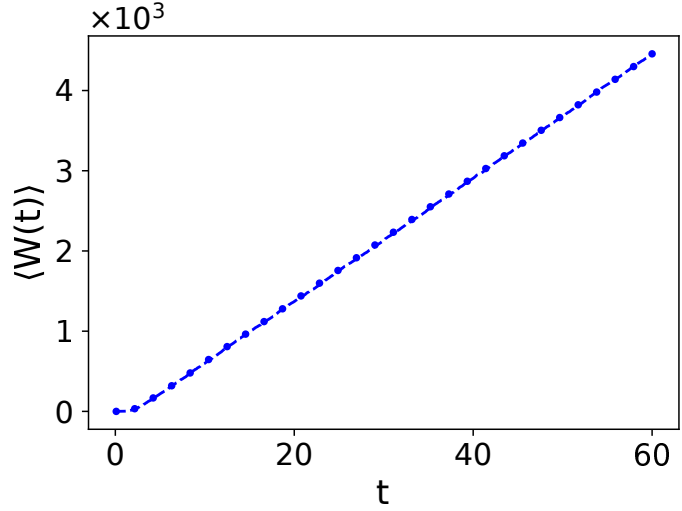}
\caption{Time evolution of the average work $\langle W(t)\rangle$: after a transient from the initial value $\langle W(0)\rangle=0$, the average work displays a linear growth with slope $\langle \dot{W}\rangle$ in Eq. \eqref{eq:dissipation}. Dashed line corresponds to simulations ($N=1000$ samples with $\Delta t=0.001$) while dots to the inverse Laplace transform of Eq. \eqref{eq:WLT}. The parameters of the model are: $D=5$, $c_+=7.5$, $c_-=-2.5$, $\nu=2.5$, $k_+=5$, $\theta_+=0.75$, $k_-=7.5$, $\theta_-=0.8$, $x_0=0$ and $\lambda=0.5$.}\label{fig:W}
\end{figure}

From Eq. \eqref{eq:dwn}, we immediately determine the first moment of the work $W(t)=\int_{\tau\in(0,t)}\mathrm{d}W(\tau)$, as
\begin{equation}\label{eq:W1t}
\begin{aligned}
\langle W(t)\rangle=&\int_{\tau\in(0,t)}\langle\mathrm{d}W(\tau)\rangle=\int_{-\infty}^{+\infty}\mathrm{d}w\,w\int_{\tau\in(0,t)}P_{\mathrm{d}W}(w|\tau)\\
=&2\kappa c_0\int_0^t\mathrm{d}\tau \sum_{\sigma_0,\sigma}\lambda_{\sigma_0}\,\sigma\left[c_m P(\sigma,\tau|\sigma_0)-\langle x_S(\sigma,\tau|x_0,\sigma_0)\rangle\right],
\end{aligned}
\end{equation}
which can be expressed in terms of the first moments $\langle x_S(\sigma,t|x_0,\sigma_0)\rangle$ of $G_S(x,\sigma,t|x_0,\sigma_0)$. Since we have access to the Laplace transform of all the parts appearing in Eq. \eqref{eq:W1t}, we take the Laplace transform of $\langle W(t)\rangle$:

\begin{equation}\label{eq:WLT}
\begin{aligned}
\langle \widetilde{W}(s)\rangle=\frac{2\kappa c_0}{s}\sum_{\sigma_0,\sigma}\lambda_{\sigma_0}\,\sigma\left[c_m \,\widetilde{P}(\sigma,s|\sigma_0)-\langle \widetilde{x}_S(\sigma,s|x_0,\sigma_0)\rangle\right], 
\end{aligned}
\end{equation}
where the expressions of $\langle \widetilde{x}_S(\sigma,t|x_0,\sigma_0)\rangle$ and $\widetilde{P}_S(\sigma,s|\sigma_0)$ are reported in Eqs. \eqref{eq:xspp}, \eqref{eq:xspm} and \eqref{eq:norms}. In Fig. \ref{fig:W} we numerically check Eq. \eqref{eq:WLT} from which the linear long-time behavior of $\langle W(t) \rangle$ clearly emerges. By considering a small-$s$ expansion of Eq. \eqref{eq:WLT}, we extract the coefficient of this linear dependence, representing the long-time average dissipated power $\langle \dot{W}\rangle$ in Eq. \eqref{eq:dissipation}.

\section{\label{app:fit} Simulation-based inference}
In order to determine the  values of the model parameters which yield the best fit to the  experimental observations we employ a simulation-based inference (SBI) toolkit implemented by \citet{greenberg2019}. Within the Bayesian framework, we define a uniform prior for the values of the parameters
$\nu \in [0.1,10]~\si{kHz}$,
$D \in [0.5,\,50]~\si{nm^2/ms}$,
$c_0 \in [1.0,\,50]~\si{nm}$,
$k \in [1.0,\,100]$,
and the period of oscillations $2 k \theta \in [10,1000]~\si{ms}$.
The SBI toolkit simulates time-series of $x(t)$ for a large sample of parameter values and thereby extracts a sample of summary statistics $\bm{\chi}[x(t)]$. These summary statistics are then used to learn the posterior distribution $p(\nu,D,c_0,k,\Theta_0|\bm{\chi})$ using the sequential neural posterior estimator \cite{greenberg2019}.

After demeaning the experimental time series, the following quantities were used as the summary statistics:
\begin{enumerate}
    \item The standard deviation $\sigma_x$ of $x(t)$;
    \item The averages $\langle{\phi_i(x/\sigma_x)}/\sqrt{\sigma_x}\rangle$ of Hermite functions $\phi_i(x)$, defined below, for $i=0,2,4$;
    \item The average, the standard deviation, and the mode of the normalized power-spectrum of $x(t)$;
    \item The Hermite-function modes $a_i$ of the autocorrelation function $\langle{x(0)x(t)}\rangle = \sum_i a_i \sqrt{\bar{f}} \,\phi_i\left(\bar{f} t\right)$, in which
    $\bar{f}$ is the average of the normalized power-spectrum of $x(t)$, for $i=0,2,4,...,12$.
\end{enumerate}
Because the probability density and the autocorrelation function of our model are even in their arguments, in the above list we employed a subset of orthonormalized even Hermite functions
$$\phi_i(z) = \mathrm{e}^{-z^2/2}
    H_i(z) \left(
            2^i i! \sqrt{\pi}
    \right)^{-1/2},$$
in which $H_i$ is an $i$-th Hermite polynomial.

\begin{figure*}[h]
\includegraphics[width = \columnwidth]{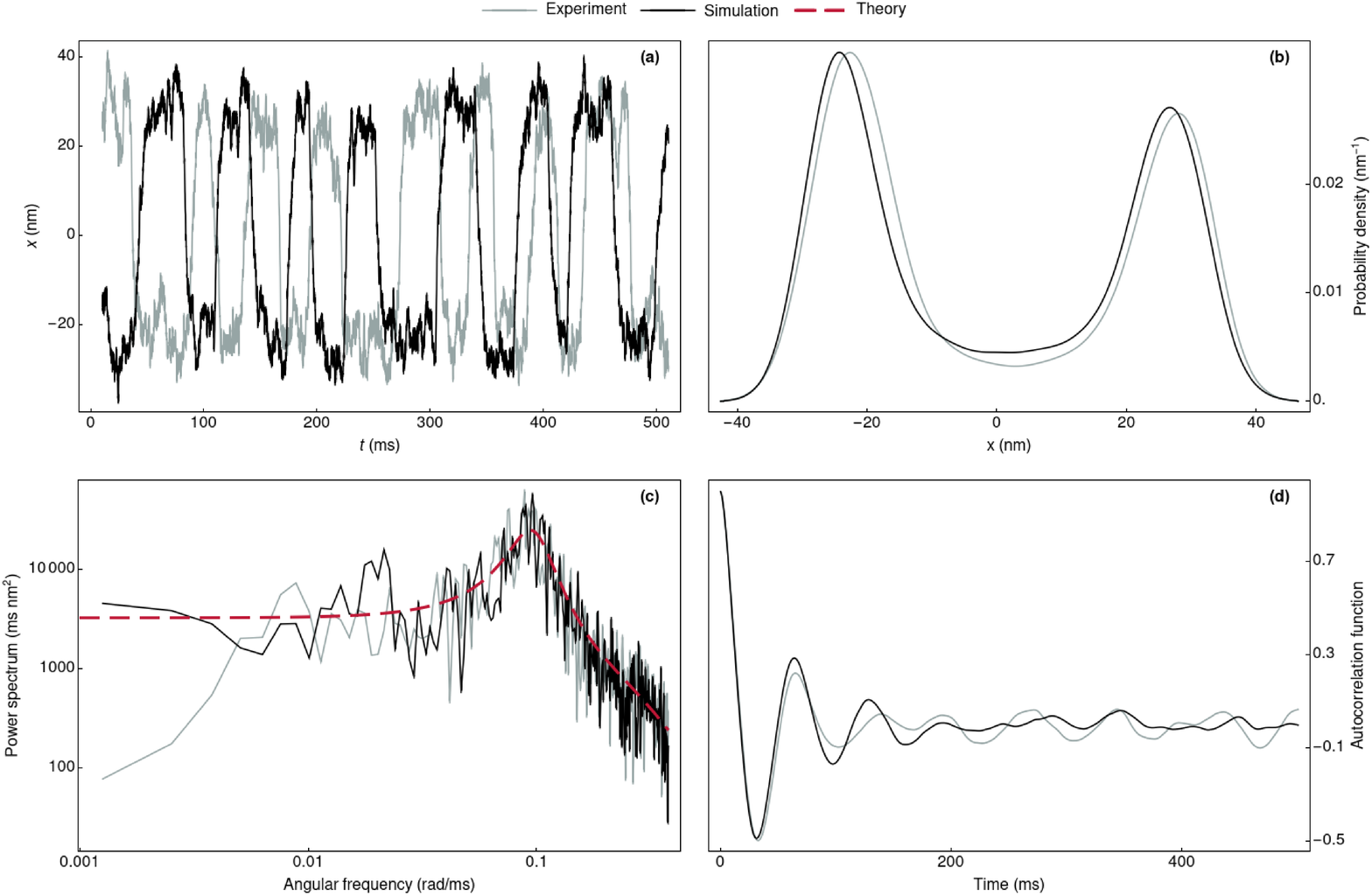}
\caption{\label{fig:run13} Case 1: fitting experimental observations of the hair-bundle oscillations to our stochastic model. Panel (a): parts of the experimental and simulated time series for the process $x(t)$. Panel (b): probability distribution densities $\rho^{\rm st}(x)$ in the experiment and simulations. Panel (c): power spectrum of the experimental and simulated time series. Panel (d): autocorrelation function for the experimental and simulated time series. The values of the parameters of the model inferred by best fitting the experimental data are:
    $\nu = \SI{0.26026\pm0.00002}{ms^{-1}}$,
    $D = \SI{5.1752\pm0.0002}{nm^2/ms}$,
    $c_0 = \SI{26.2755\pm0.0008}{nm}$,
    $k=7.133\pm0.002$,
    $\theta=\SI{4.542\pm0.002}{ms}$.
}
\end{figure*}

\begin{figure*}[h]
\includegraphics[width = \columnwidth]{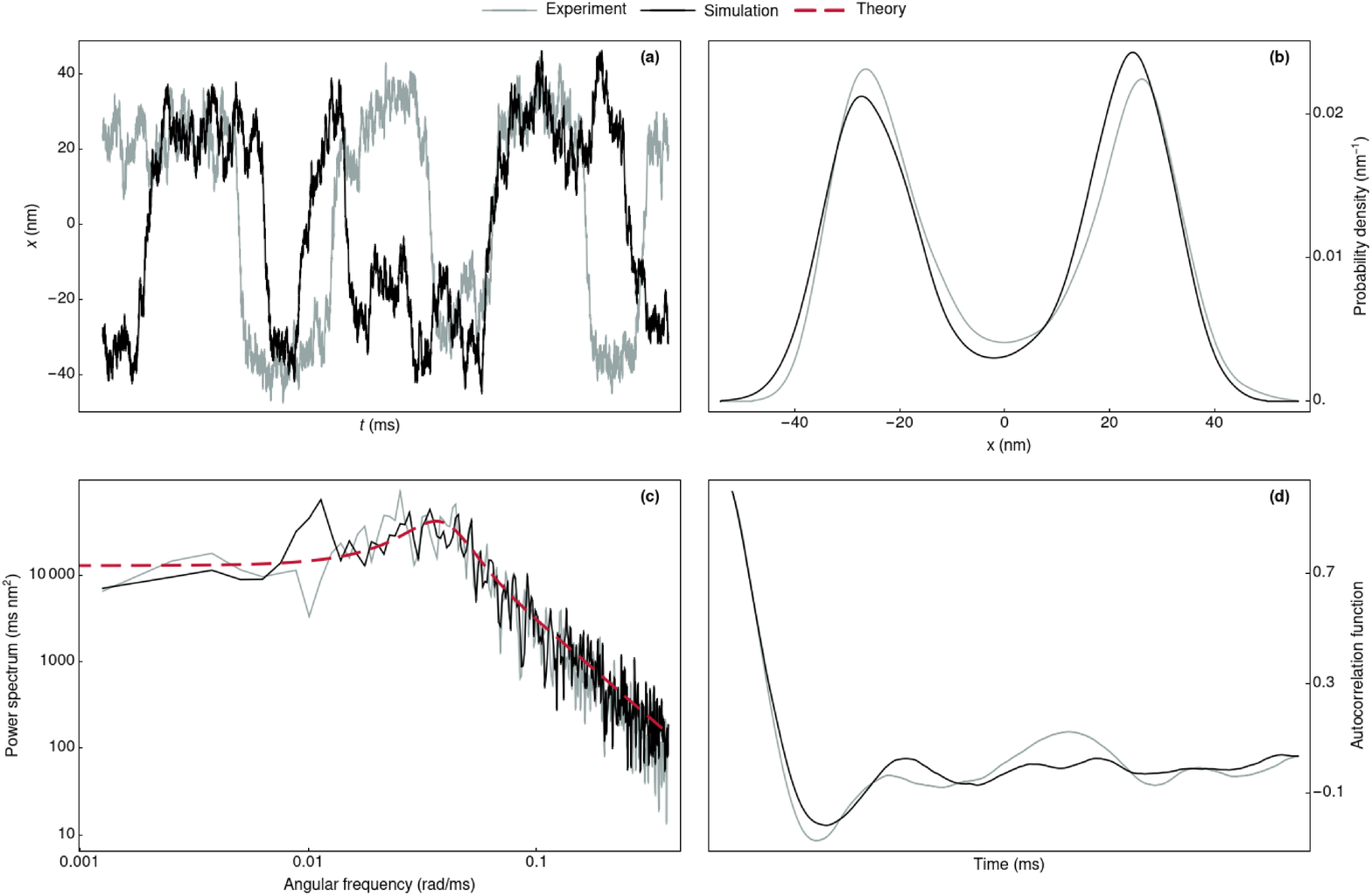}
\caption{\label{fig:tra8} Case 2: fitting experimental observations of the hair-bundle oscillations to our stochastic model. Panel (a): parts of the experimental and simulated time series for the process $x(t)$. Panel (b): probability distribution densities $\rho^{\rm st}(x)$ in the experiment and simulations. Panel (c): power spectrum of the experimental and simulated time series. Panel (d): autocorrelation function for the experimental and simulated time series. The values of the parameters of the model inferred by best fitting the experimental data are:
    $\nu = \SI{0.172\pm0.002}{ms^{-1}}$,
    $D = \SI{9.180\pm0.003}{nm^2/ms}$,
    $c_0 = \SI{25.991\pm0.002}{nm}$,
    $k=4.267\pm0.006$,
    $\theta=\SI{18.40\pm0.04}{ms}$.
}
\end{figure*}

\begin{figure*}[h]
\includegraphics[width = \columnwidth]{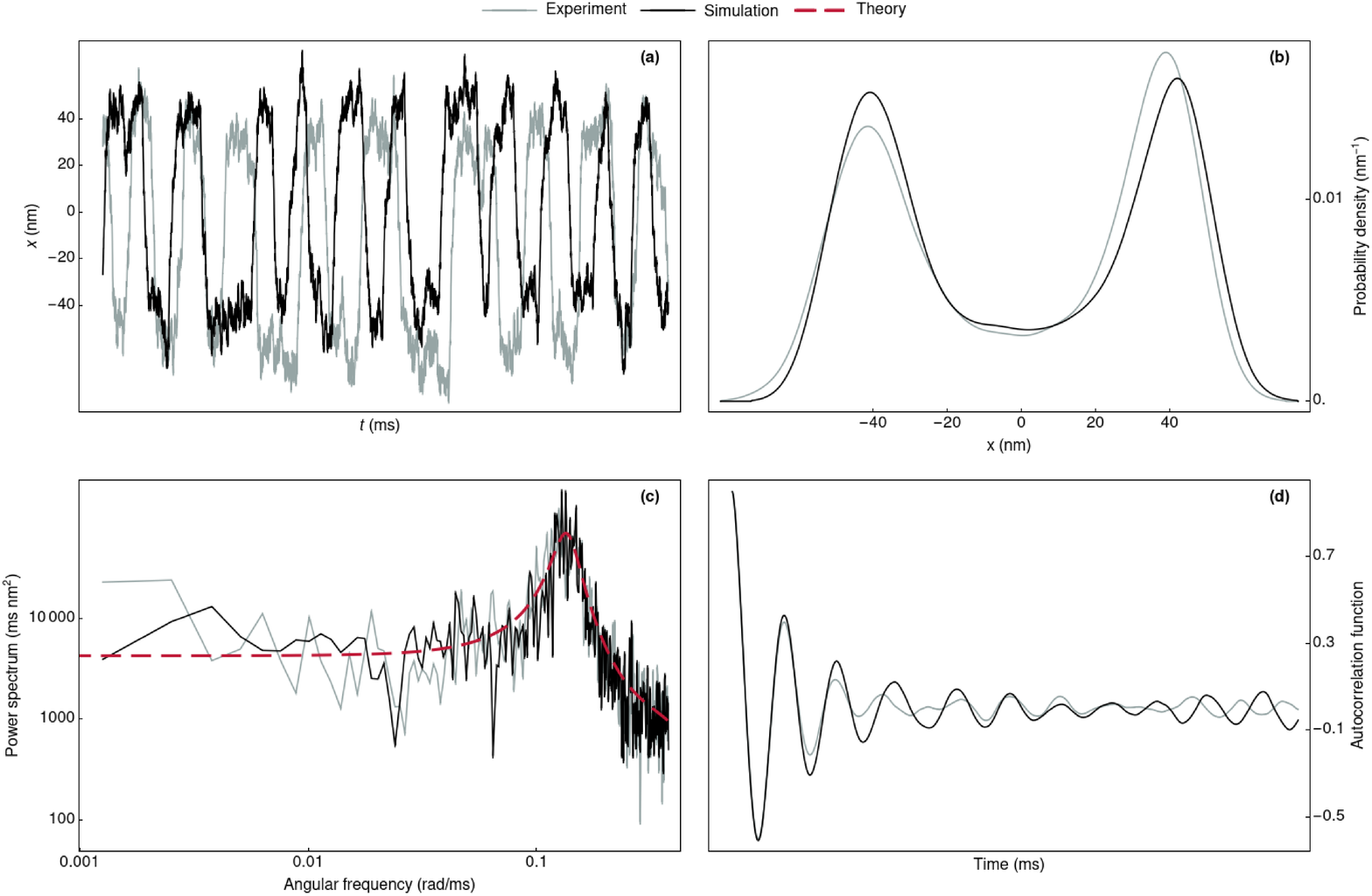}
\caption{\label{fig:f1116} Case 3:  fitting experimental observations of the hair-bundle oscillations to our stochastic model. Panel (a): parts of the experimental and simulated time series for the process $x(t)$. Panel (b): probability distribution densities $\rho^{\rm st}(x)$ in the experiment and simulations. Panel (c): power spectrum of the experimental and simulated time series. Panel (d): autocorrelation function for the experimental and simulated time series. The values of the parameters of the model inferred by best fitting the experimental data are:
    $\nu = \SI{0.28136\pm0.00002}{ms^{-1}}$,
    $D = \SI{18.7724\pm0.0008}{nm^2/ms}$,
    $c_0 = \SI{43.850\pm0.001}{nm}$,
    $k=11.591\pm0.003$,
    $\theta=\SI{1.9979\pm0.0006}{ms}$.
}
\end{figure*}

Three inference rounds of $30\times10^3$ simulations were performed to learn the posterior distribution from summary statistics. We considered only 10-\si{s}-long time series sampled with a time step of \SI{0.1}{ms}. By fixing the summary statistics to experimental observations, the posterior distribution was then sampled to obtain the model parameter values and their uncertainties. We applied this procedure to the three experimental time series reported in Figs.~\ref{fig:run13}--\ref{fig:f1116}, in which we found symmetric oscillations of the hair-bundle tip.
Finally, we report in Table \ref{tab:1} the estimate of the stationary average power per cycle expressed by Eq. \eqref{eq:dissipation} for the parameters inferred for Figs. Figs.~\ref{fig:run13}--\ref{fig:f1116}. 

\begin{center}
\begin{table}
\centering
  \begin{tabular}{ | c | c | c | c |}
    \hline
    Case & $\langle \dot{W}\rangle\,\,(k_B T/\text{cycle})$ & Error bar\\
    \hline
    1 & 129.28 & 0.05\\ \hline
    2 & 50.4 & 0.6\\ \hline
    3 & 85.37 & 0.07\\ \hline
  \end{tabular}
\caption{Estimate of the stationary average power per cycle $\langle\tau\rangle \langle\dot{W}\rangle$ in units of $k_B T$ via Eq. \eqref{eq:dissipation} for the parameters in Figs. \ref{fig:run13}, \ref{fig:tra8}, and \ref{fig:f1116}. }
\label{tab:1}
\end{table}
\end{center}

\end{document}